\documentclass[iop]{emulateapj}

\usepackage{amssymb}
\usepackage{times}
\usepackage{graphicx}
\usepackage{url}
\usepackage{textcomp}
\usepackage{calc}
\usepackage{chngcntr}

\newcommand{\nbody}{$N$-body}
\newcommand{\msun}{M_{\sun}}
\newcommand{\mcl}{M_{\mathrm{ecl}}}
\newcommand{\mclmax}{M_{\mathrm{ecl,max}}}
\newcommand{\mclmin}{M_{\mathrm{ecl,min}}}
\newcommand{\mmax}{m_{\mathrm{max}}}
\newcommand{\dm}{\mathrm{d}m}
\newcommand{\comp}{{}^{\mathrm{com}}}
\newcommand{\suth}{{}^{\mathrm{th}}}
\newcommand{\tot}{{}^{\mathrm{total}}}
\newcommand{\msys}{m_{\mathrm{sys}}}
\newcommand{\nej}{N_{\mathrm{ej,O}}}
\newcommand{\fej}{f_{\mathrm{ej,O}}}
\newcommand{\nos}{N_{\mathrm{O}}}
\newcommand{\rh}{r_{\mathrm{h}}}
\newcommand{\kms}{\ensuremath{\mathrm{km~s^{-1}}}}
\newcommand{\uden}{\ensuremath{\msun\mathrm{pc}^{-3}}}
\newcommand{\afej}{\langle \comp \fej \rangle}
\newcommand{\anej}{\langle \comp \nej \rangle}
\newcommand{\anos}{\langle \comp \nos \rangle}
\newcommand{\vcl}{\sigma_{\mathrm{cl}}}

\shorttitle{O star ejection as a function of $\mcl$}
\shortauthors{Oh, Kroupa \& Pflamm-Altenburg}

\begin{document}

\title{Dependency of dynamical ejections of O stars on the masses of very young star clusters}

\author{Seungkyung Oh, Pavel Kroupa and Jan Pflamm-Altenburg} 
\affil{Helmholtz-Institut f\"ur Strahlen- und Kernphysik (HISKP), University of Bonn, 
Nussallee 14-16, D-53115 Bonn, Germany}
\email{skoh@astro.uni-bonn.de}

\begin{abstract}
Massive stars can be efficiently ejected from their birth clusters
through encounters with other massive stars. 
We study how the dynamical ejection fraction of O star systems varies with 
 the masses of very young star clusters, $\mcl$, by means of direct \nbody\ calculations.
We include diverse initial conditions by varying the half-mass radius, 
initial mass-segregation, initial binary fraction and orbital parameters of the massive binaries. 
The results show robustly that the ejection fraction of O star systems exhibits a 
maximum at a cluster mass of $10^{3.5}~\msun$ for all models, 
even though the number of the ejected systems increases with cluster mass.   
We show that lower mass clusters ($\mcl\approx 400~\msun$) are the dominant sources for populating 
the Galactic field with O stars by dynamical ejections, considering the mass function of embedded clusters.
About 15 per cent (up to $\approx$38 per cent, depending on the cluster models) 
of O stars of which a significant fraction are binaries, and 
which would have formed in a $\approx$10~Myr epoch of star formation 
in a distribution of embedded clusters, will be dynamically ejected to the field. 
Individual clusters may eject 100 per cent of their original O star content. 
A large fraction of such O stars have velocities up to only $10~\kms$.
 Synthesising a young star cluster mass function it follows, 
given the stellar-dynamical results presented here, that the observed fractions of field 
and runaway O stars, and the binary fractions among them can be well understood theoretically 
if all O stars form in embedded clusters. 
\end{abstract}


\section{Introduction}
The majority of O stars are found in young star clusters or OB associations. 
It has been suggested that a small fraction of massive stars may form in isolation 
as a result of stochastic sampling of the stellar and cluster mass function \citep[see][]{Wet05,PG07,OL12}.  
However, in the case of the Galactic field O stars within 2--3~kpc from the Sun,  
essentially all of them can be traced back to young star clusters or associations \citep{SR08}, 
suggesting the field O stars also formed in clusters \citep{Get12}.
Some observational studies on O stars in the Large and Small Magellanic Clouds  
have claimed there to be evidence for isolated massive star formation \citep{Bet12,Oet13}, 
such as e.g., apparent isolation, absence of a bow-shock, and circular shape of an \ion{H}{2} region. 
These claims can not, however, be conclusive since they can be explained with slowly moving runaways 
or former members of clusters which have already dissolved \citep{Weidner07}. 
For example, only 30--40 per cent of the Galactic OB runaways are found to have a bow-shock \citep{VND95,HK02}, 
and, furthermore, bow-shock formation depends on the physical condition of the ambient medium 
through which a runaway travels \citep[e.g., temperature and density of the ambient medium,][]{HK02}.
\citet{MLV13} showed that runaways can have a circular \ion{H}{2} region in the projected H$\alpha$ emission. 
Furthermore, in the case of the 30~Dor region Bressert et al. considered, 
there are many young star clusters in the region. If the 30~Dor region is modelled by a population 
of clusters, how many O stars would be ejected or removed from their clusters in 10 Myr? 
This has not been investigated. Any conclusion on the issue of isolated O-star formation 
are, for the time being, thus premature. 
Finally, measuring stellar proper motions in the Large or Small Magellanic Clouds is unfeasible such that claims for O stars formed in isolation become next to impossible to confirm or reject. 
 With \facility{Gaia}, however, it may become possible to put constraints on this 
if proper motions of massive stars can be measured to a precision of a few $\kms$ in the Magellanic Clouds.  
Given the above results on the well studied ensemble of O stars within
2--3~kpc around the Sun it appears physically more plausible that all O stars form in embedded clusters. 
 
There are two mechanisms to expel O stars from a star cluster. 
The one is the binary supernova scenario \citep{Bl61,PS00,ELT11} in which a star in a binary obtains a high velocity 
after the supernova explosion of the initially more massive companion star. The other is dynamical ejection through strong close encounters with other massive stars in the cluster core \citep{PRA67,LD90,FP11,BKO12,PS12}. 
Of the candidate isolated O stars catalogued by \citet{SR08}, 
$8$--$9$ per cent could not be traced back to any still existing star clusters or associations, 
 but are consistent with originating in a cluster which could not be found due to several
effects such as incompleteness of their cluster sample \citep{SR08}, dissolution of the cluster \citep{Get12}, and having experienced the two-step-ejection process proposed by \citet{PK10}. 
The majority of the stars which can be traced back to clusters appear to have been ejected when their parental clusters were very young \citep{SR08}.
This may suggest that dynamical ejections through few-body interactions in the heavy-star-rich cluster core mainly populate the field O star population \citep{CP92}. 

There have been many studies on dynamically ejected massive stars from young star clusters 
using numerical integrations. In most cases the studies performed few-body scattering experiments 
focused on runaways, which are ejected with a velocity higher than $\approx$30--40~$\kms$   
\citep{GPE04,PK06,GGP09,GG11}. 
Full \nbody\ calculations for a particular cluster mass have also been presented \citep*{BKO12,PS12}. 
\citet{FP11} performed \nbody\ calculations for very dense star clusters composed of single stars
with three different cluster masses ($\geq 6000~\msun$) and 
showed that the clusters with the lowest mass in their models, $6000~\msun$, 
produce runaways most efficiently. They argued that the one bullying binary formed 
during core collapse of the cluster is responsible for the production of runaways 
and that the number of runaways is almost independent of cluster mass. 
The runaway fraction (number of runaway stars over total number of stars still present) 
therefore decreases with cluster mass as the more massive clusters have more stars. 
Note that the lower-mass limit for stars in their models is $1~\msun$. 
Thus, if a cluster is populated with a full range of stellar masses from the hydrogen burning 
limit upwards, in order to have the same amount of massive stars,  
the cluster would be twice as massive than they claimed. 
The relation between O star ejection fraction and cluster mass has not been studied in detail to date. 

Here we study for the first time how the dynamical ejection fraction of O stars  
changes with the masses of very young star clusters using direct \nbody\ calculations 
with a broad range of cluster masses, from low-mass clusters containing one or two O stars 
to massive clusters containing a few hundred O stars, and including primordial binaries. 
Such an investigation based on a very large library of young cluster models 
has not been possible to date because the knowledge on the properties of initial stellar 
population, their birth configurations and last but not least the computational algorithms 
enabling such a CPU-intensive and massive numerical challenge have not been in place. 
Thus, for  example, how to initialize initially a mass segregated cluster in dynamical equilibrium is 
only known since the work of \citet{SKB08} and \citet{BDK08}  
and the statistical distribution functions of massive star binaries are becoming 
observationally constrained only now \citep[e.g.,][]{Set12,Set13,KK12,Ket14,Aet15}. 
The \nbody\ calculations, made possible by the relentless code and algorithmic advances 
by Sverre Aarseth and Seppo Mikkola, used in this study are briefly described in Section~2.
The results are presented in Sections~3--6, and the discussion and summary follow in Sections~7 and 8, respectively.

\section{{\it N}-body models}
The data used in this study are based on the theoretical young star cluster library 
of model sequences over cluster mass computed by \citet{OK12} with direct \nbody\ calculations using the NBODY6\footnote{The code can be freely downloaded from 
\url{http://www.ast.cam.ac.uk/\texttildelow{}sverre/web/pages/nbody.htm}.} 
code \citep{Aa03} and various initial conditions. 
Among the model sequences in the library, we adopt four sequences here and, further, 
extend them to higher cluster masses. 
Additionally we perform calculations for two more sets of initial conditions for this study. 
Table~\ref{tabIC} lists properties of the model sequences studied in this paper. 
The initial setup of the models is described below. 

\begin{deluxetable}{lccccc}
\tablewidth{0pt}
\tablecaption{List of \nbody\ model sequences studied here.\label{tabIC}}
\tablehead{
\colhead{Name} & \colhead{$\rh$ (pc)} & \colhead{IMS} &\colhead{IBF}&\colhead{IPD}&\colhead{Pairing method}
}
\startdata
MS3OP         &0.3      &Yes  &1  &Kroupa95      & OP\\
NMS3OP        &0.3      &No   &1  &Kroupa95      & OP\\
MS3OP\_SP     &0.3      &Yes  &1  &Sana et al.12 & OP\\
MS3UQ\_SP     &0.3      &Yes  &1  &Sana et al.12 & uniform $q$ dist.\\
MS3S          &0.3      &Yes  &0  &\nodata       & \nodata\\
MS8OP         &0.8      &Yes  &1  &Kroupa95      & OP\\
MS1OP         &0.1      &Yes  &1  &Kroupa95      & OP
\enddata
\tablecomments{Names of model sequences and initial half-mass radii ($\rh$) are listed in columns 1 and 2, respectively. 
 Columns 3--6 present initial mass segregation (IMS), initial binary fraction (IBF), and 
 initial period distribution (IPD) and pairing method used for massive binaries (primary mass $\ge 5~\msun$).}
\end{deluxetable}

The four model sequences that we adopt from the library for this study are comprised of 
binary-rich clusters with initial mass segregation (MS3OP) and without initial mass segregation (NMS3OP), 
single-star clusters (MS3S) with an initial half-mass radius, $\rh$, of 0.3~pc, 
and initially mass-segregated, binary-rich clusters with an initial $\rh$ of 0.8~pc (MS8OP). 

The cluster masses, $\mcl$, in the library range from $10$ to $10^{3.5}~\msun$ (to $10^{4}~\msun$  
for the MS3OP model sequence). For extending the range of cluster masses,
we additionally carried out calculations for clusters with $\mcl \approx 10^{4}$ and $10^{4.5}~\msun$
($\mcl \approx 10^{4.5}~\msun$ for the MS3OP model sequence). 
Individual stellar masses are randomly drawn from the canonical initial mass function (IMF), 
which is a two-part power-law \citep{KP01,Ket13},
\begin{equation}
 \xi(m) =k \left\{ \begin{array}{ll}
 \left(\frac{m}{0.08} \right)^{-1.3}, & 0.08 \le m/\msun < 0.50, \\
  \left(\frac{0.5}{0.08} \right)^{-1.3} \left( \frac{m}{0.5} \right)^{-2.3},
 & 0.50 \le m/\msun \le \mmax.
 \end{array} \right.
\label{imf}
\end{equation}
 More details of our stellar mass sampling can be found in \citet{OK12}.
The mass of the most massive star in the cluster, $\mmax$, is chosen from the maximum-stellar-mass -- 
cluster-mass relation \citep{WK04,WKB10,WKP13,WKP14}. Thus only clusters with $\mcl \geq 10^{2.5}~\msun$ 
initially have O-type stars in our models. \citet{PWK07} provide a fitting formula for $\mmax(\mcl)$. 
This procedure was adopted rather than optimal sampling introduced in \citet{Ket13} 
because the theoretical young cluster library of \citet{OK12} did not have this sampling method to disposal.

\begin{figure}
 \center
 \includegraphics[width=75mm]{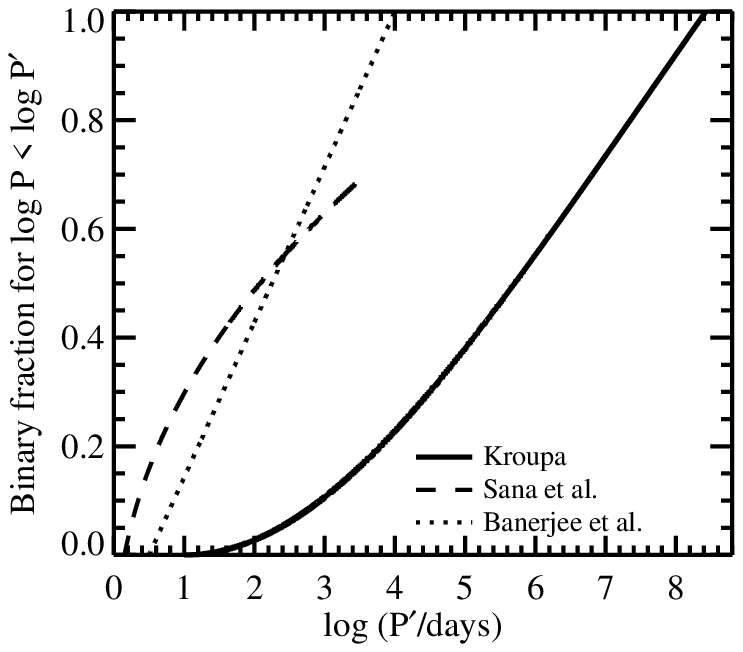}
 \caption{Cumulative binary fraction for $\log_{10}P < \log_{10}P'$ among all O stars.
 Solid and dashed lines are calculated from the \citet[Equation~(\ref{KP})]{PK95b} 
 and \citet[Equation~(\ref{SP})]{Set12} period distributions, respectively.
 Dotted line is a uniform period distribution with a period range of $0.5 < \log_{10}(P/\mathrm{days}) <4$
 used in \citet{BKO12} for O star binaries. For example, 49 per cent of all O stars are binaries with
 period shorter than 100~d according to the Sana et al. distribution.
 }
 \label{pcdf}
\end{figure}

Binary-rich models have an initial binary fraction of 100 per cent, i.e., all stars are in 
a binary system at $t=0$~Myr.
This is motivated by the angular momentum problem of star formation since the angular momentum 
of a collapsing cloud core can be distributed efficiently into two stars, and also 
by the strong empirical evidence that star formation universally prefers the binary mode  \citep{Det07,GK05,Get07,MK12,Leigh15,Set14}. 
The \citet{PK95b} period distribution function with minimum and maximum periods of
10 and $10^{8.43}$~days, respectively, is adopted for all binaries in the library.
The Kroupa period distribution, 
\begin{equation}
f_{P} = 2.5{{{\log_{10} P - 1 }} \over {45 + (\log_{10} P - 1)^{2}}},
\label{KP}
\end{equation}
where the period ($P$) is in days, has been derived iteratively using binary-star data 
from the Galactic field population and from star forming regions. 
With this distribution, about 2.75 per cent of all O star binaries have a period shorter than 100~days (Figure~\ref{pcdf}).
Although it was derived from binaries with primary star mass $\leq 1~\msun$, 
the same distribution is used for more massive binaries in the library since no
well constrained period distribution of massive binary population was available
when the \nbody\ calculations were begun.

However, recently, \citet{Set12} derived an intrinsic period distribution of O stars of the form
$f_{P} \propto (\log_{10}P) ^{-0.55}$,
over the period range of $0.15 < \log_{10} (P/{\mathrm{days}}) < 3.5$, from spectroscopic observations
of O-star populations in nearby young open star clusters. The \citet{Set12} period distribution
leads to $\approx49$ per cent of O stars being binaries with a period shorter than 100~d (Figure~\ref{pcdf}).
Here we additionally perform a set of \nbody\ integrations with massive binaries ($m\ge5~\msun$)
having the \citet{Set12} period distribution, MS3OP\_SP, but with the maximum period extended to the value
$\log_{10} (P/{\mathrm{days}})\approx 6.7$ at which the cumulative binary fraction becomes unity.
We thus adopt the following period distribution function for the massive binaries,
\begin{equation}
f_{P} = 0.23\times(\log_{10}P)^{-0.55},
\label{SP}
\end{equation}
for which $\int^{6.7}_{0.15}f_{P} \, d\log_{10}P = 1$.
The other initial conditions are the same as those of MS3OP. 
See also \citet{KK12}, \citet{Ket14} and \citet{Aet15} 
for slightly different period distributions of O-star binaries derived from observations.

Stars with a mass $m < 5~\msun$ are randomly paired after choosing the two masses randomly 
from the IMF. Again, this mass-ratio distribution is derived from the Galactic field population 
and from star forming regions \citep{PK95a,PK95b}. 
Stars with $m\ge5~\msun$ are paired to imprint observations that the massive binaries favour 
massive companions \citep{PS06,KF07,KK12,Set12}.
To achieve this we first order all stars more massive than $5~\msun$ in to an array of decreasing mass. 
The most massive star is then paired with the next massive star, the third most massive star is paired with 
the fourth most massive star, and so on. We refer to this procedure as ordered pairing (OP). 
Whilst with this method it is easy to have a massive primary being paired with a massive companion 
while preserving the IMF, it gives a significant bias toward high mass ratios
($q\approx1$ where $q=m_{2}/m_{1}$ and $m_{1}\ge m_{2}$) which is not supported by recent observations \citep{KK12,Set12,Ket14}. The recent observations show the mass-ratio distribution of O-type binaries 
to be rather uniform, 
\begin{displaymath}
f(q) \propto q^\eta
\end{displaymath} 
with $-0.1\lesssim \eta \lesssim0.1$ and $0.1 \leq q \leq 1.0$. 
We add a model sequence (MS3UQ\_SP) with a uniform mass-ratio distribution, $\eta = 0$, 
for massive binaries ($m_{1}\ge 5~\msun$), being otherwise the same as the MS3OP\_SP model sequence. 
First, a mass-ratio value for a primary is derived from a uniform probability function and 
then the secondary is chosen from the rest of the members which gives the closest $q$ 
value for the derived one. This is important to preserve the IMF: It is important to first 
generate a full list of stars whose combined mass corresponds to the mass of the cluster and only then 
to create the binaries using this list only. 
The thermal distribution is used for the initial eccentricity distribution, i.e., $f_{e}(e)=2e$, 
where $e$ is the eccentricity of a binary \citep[see][]{Kr08}.

Each model sequence has one initial half-mass radius regardless of the cluster mass  
since by observation there is no significant relation between the size and mass of a cluster \citep{Bet05,Schet07} 
or at most a only very weak one \citep{LS04,MK12}. 
The cluster-mass--initial half-mass radius relation from \citet{MK12}, 
\begin{displaymath}
 (\rh/\mathrm{pc})=0.1\times (\mcl/\msun)^{0.13},
\end{displaymath} 
gives initial half-mass radii of $\approx0.21$ to $0.38$~pc for $\mcl = 10^{2.5}$--$10^{4.5}~\msun$.
Our main, initial half-mass radius (0.3~pc) lies close to the median value and 
is well consistent with the typical initial half-mass radius (0.1--0.3~pc) inferred for the Milky Way clusters \citep{MK11}.   
Thus, our results may be able to represent reality. 
Here we include two more model sequences with different half-mass radii of 0.1 and 0.8 pc, 
 MS1OP and MS8OP, respectively. This allows us to estimate how the ejection fraction changes with cluster mass 
when the cluster radius varies with cluster mass. 
 While the MS8OP model sequence is adopted from the library of \citet{OK12}, 
the MS1OP model sequence is computed for this work. 
From the choices of half-mass radii, the initial central densities of our models range from $\approx 300$ 
to $3.3\times10^{4}~\uden$ for $\rh=0.8$~pc clusters, from $6.2\times10^{3}$ to $6.2\times10^{5}~\uden$ 
for $\rh=0.3$~pc clusters, and from $\approx1.7\times10^{5}~\uden$ to $\approx1.7\times10^{7}~\uden$ 
for $\rh=0.1$~pc clusters.  

For all models, initial positions and velocities of single stars or of centre-of-masses of binaries 
in the clusters are generated following the Plummer density profile under the assumption that 
the clusters are in virial equilibrium.
Initially mass-segregated clusters are produced being fully mass-segregated with the method 
developed in \citet*{BDK08} in which more massive systems are more bound to the cluster.

The \nbody\ code regularises the motions of stars in compact configurations and
incorporates a near-neighbour force evaluation scheme to guarantee the high accuracy of the solutions
of the equations of motions. Stellar evolution is taken into account by using the stellar evolution 
library \citep{HPT00,HTP02} implemented in the code.  All clusters are evolved up to 3~Myr, 
i.e., to a time before the first supernova occurs, in order to study dynamical ejections only.  
Moreover the dynamical ejections of O stars are most efficient at an early age of the cluster
(Oh et al. in preparation and see Section~5), 
thus computations to an older cluster age would not change our conclusion on the dynamical ejections. 

A standard solar-neighbourhood tidal field is adopted. The clusters are highly tidally underfilling 
such that the tidal field of the hosting galaxy plays no significant role for the results presented here. 
Our cluster models are initially gas-free, i.e., without a background gas-potential,
and thus the gas-expulsion phase is not included here. 
While 100 realisations with different random seed numbers are performed
for the clusters with $\mcl \leq 10^{3.5}~\msun$, 10 and 4 realisations are carried out 
for $\mcl=10^{4}$ and $10^{4.5}~\msun$ clusters, respectively, as computational costs
increase with the number of stars, i.e., cluster mass, and stochastic effects
are reduced with a larger number of stars.

\section{Ejection fraction of O stars as a function of $\mcl$}
\begin{figure*}
 \includegraphics[width=59mm]{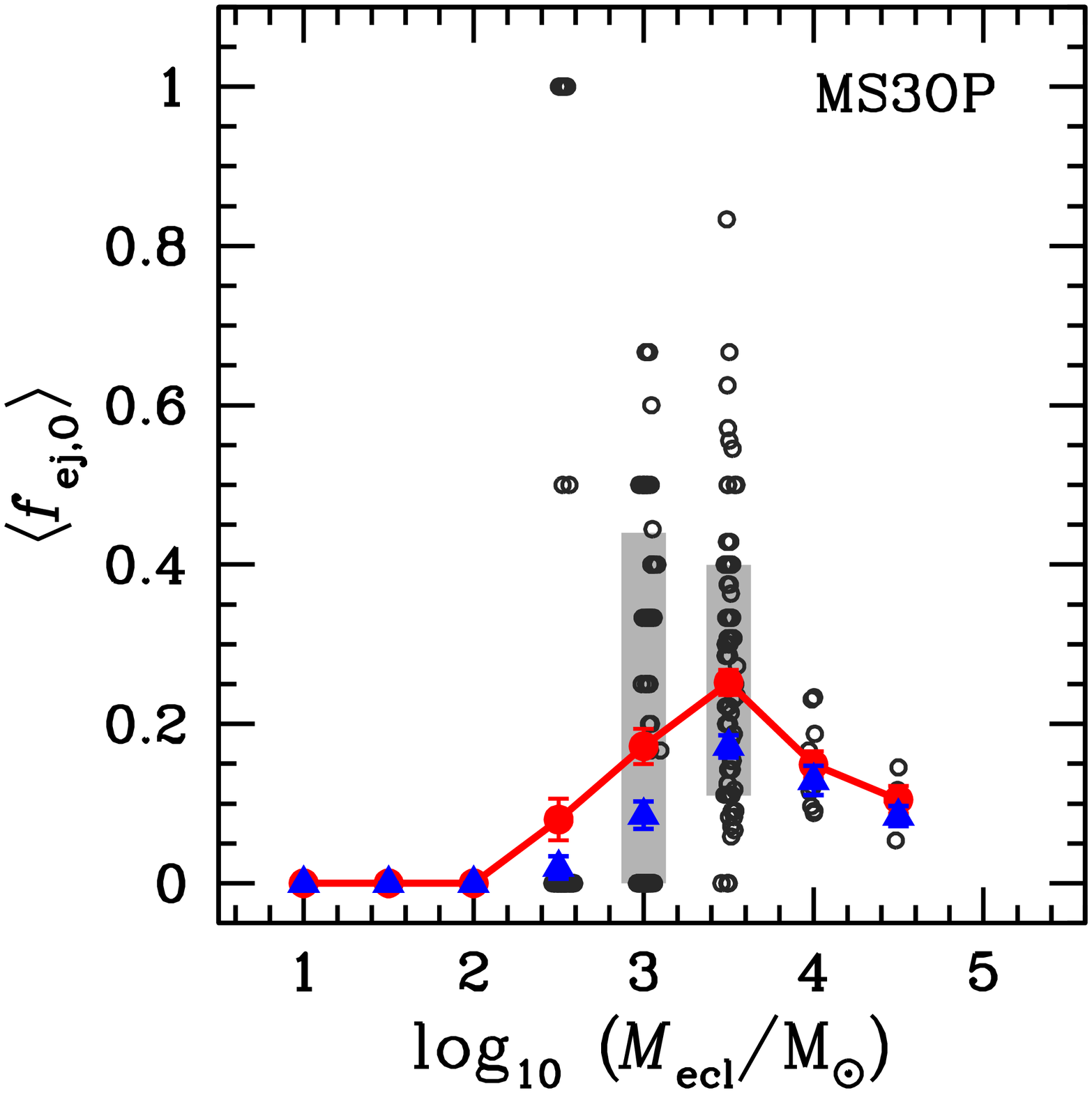}
 \includegraphics[width=59mm]{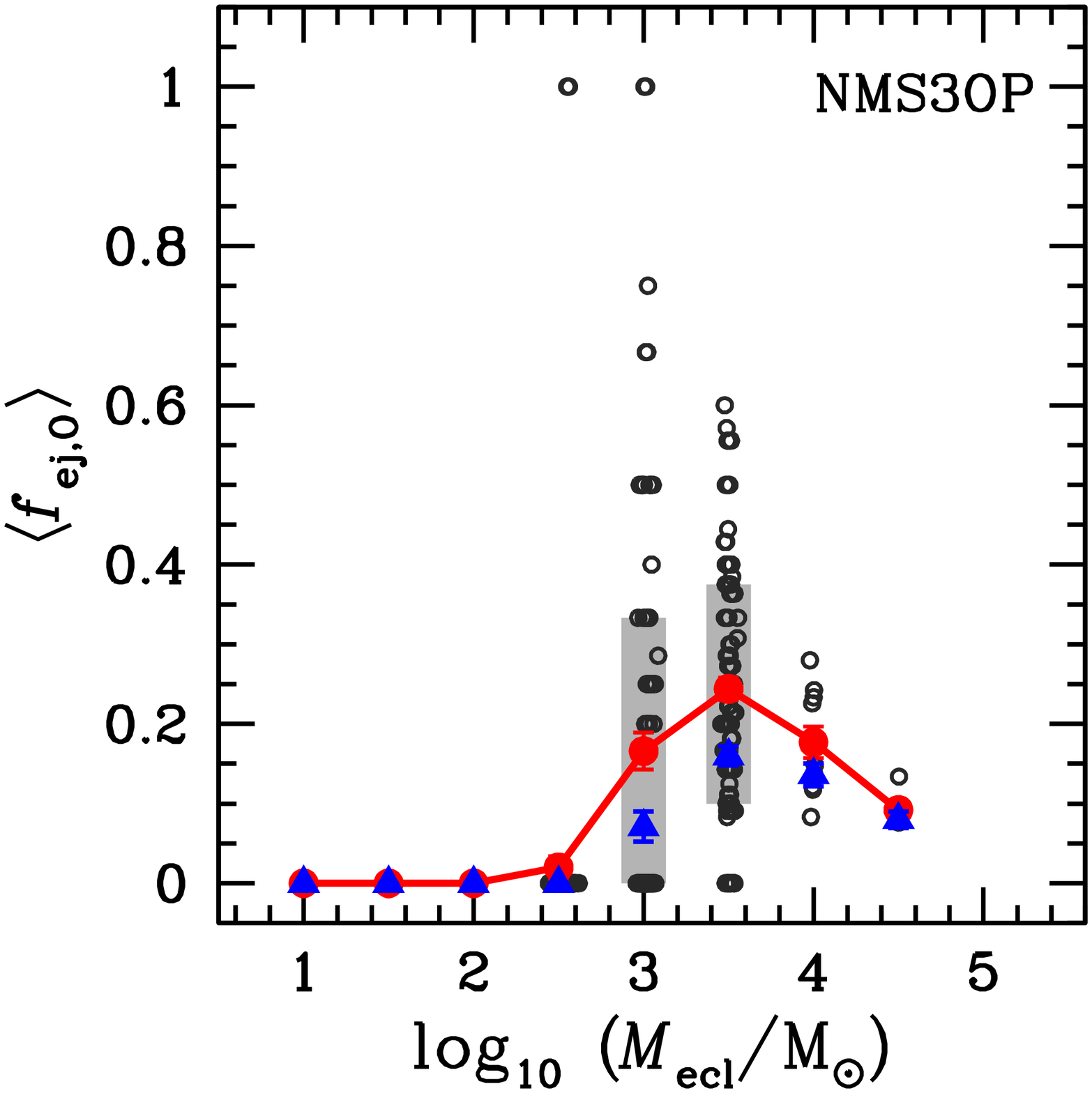}
 \includegraphics[width=59mm]{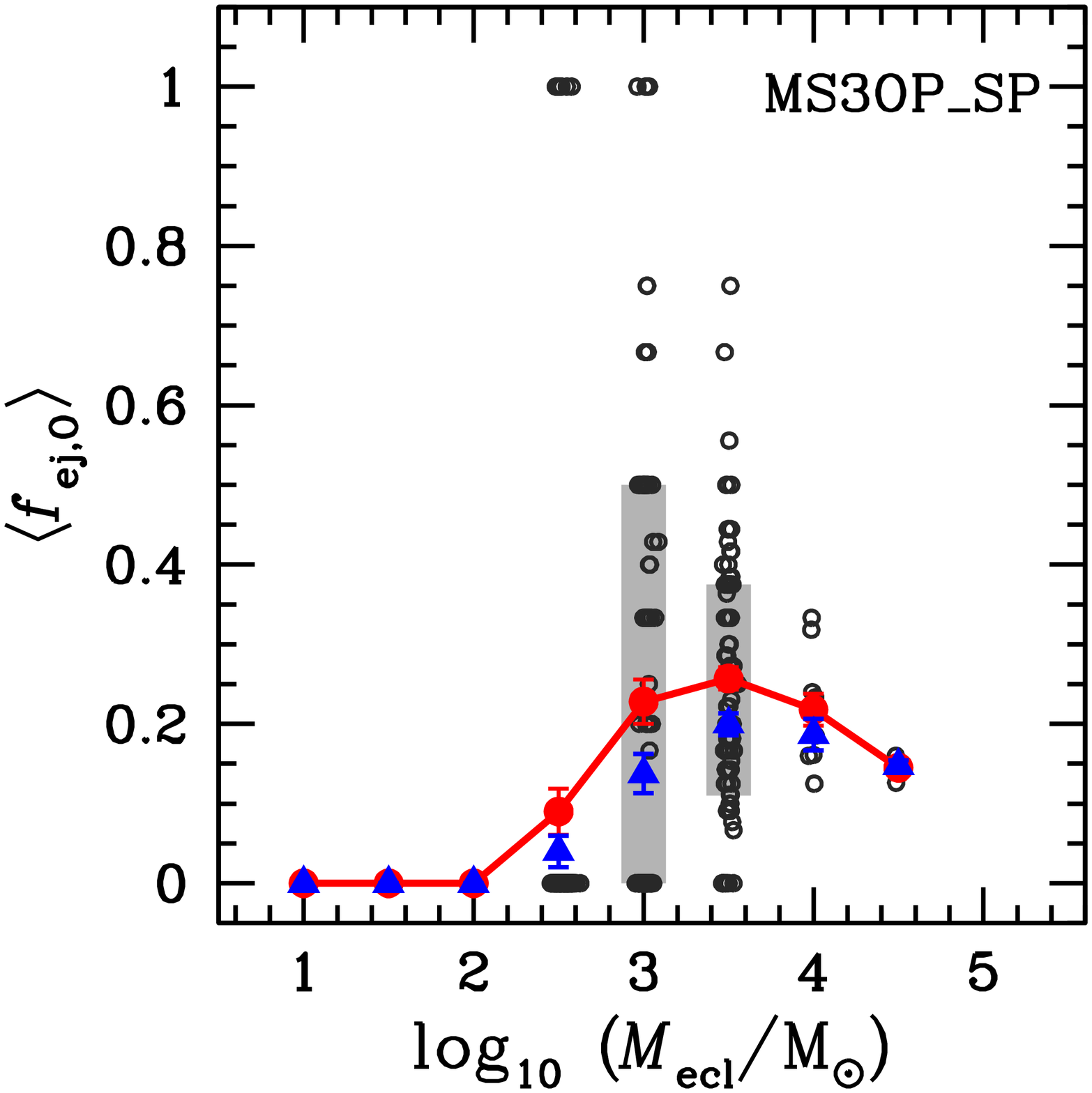}\\
 \includegraphics[width=59mm]{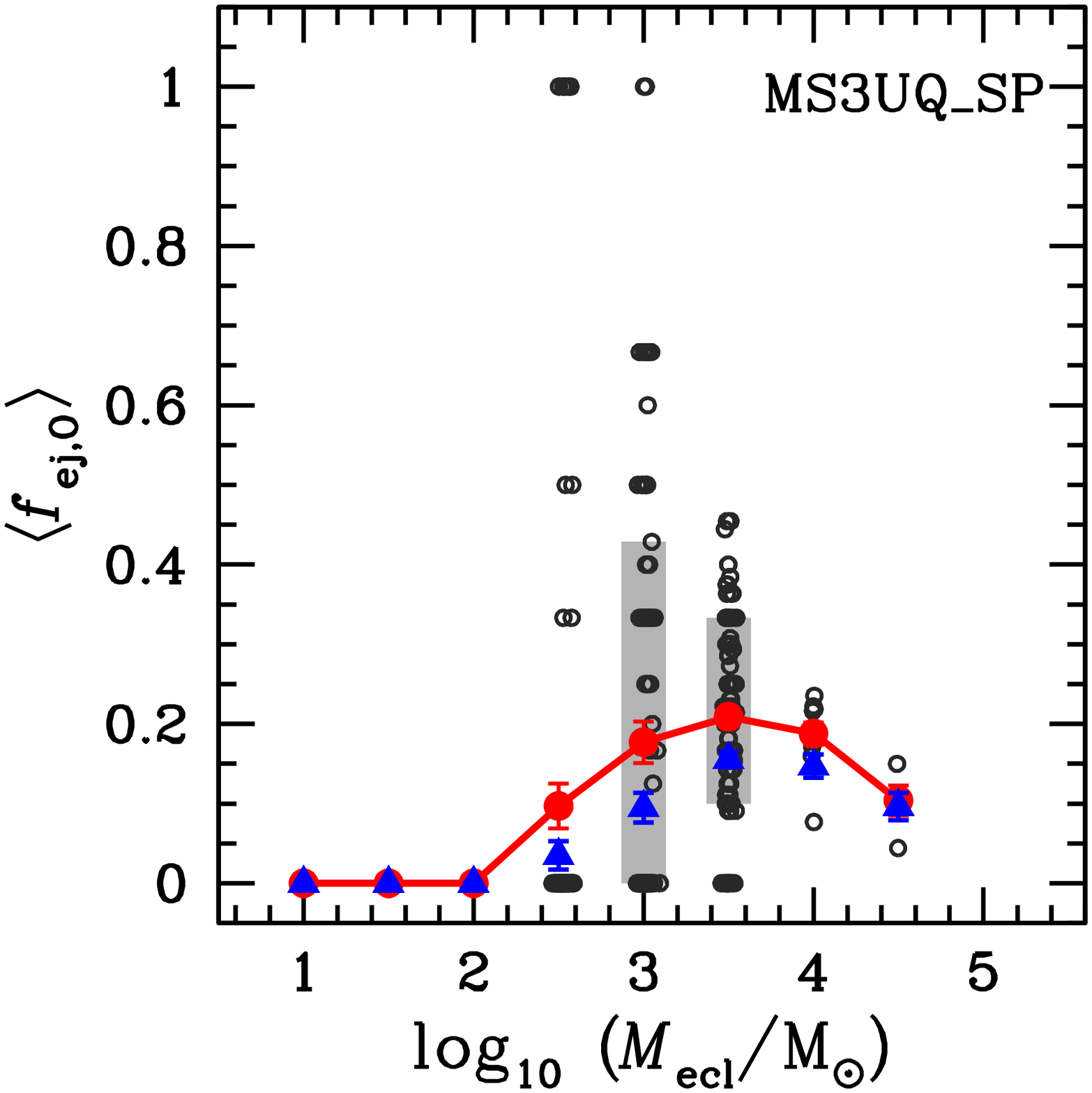}
 \includegraphics[width=59mm]{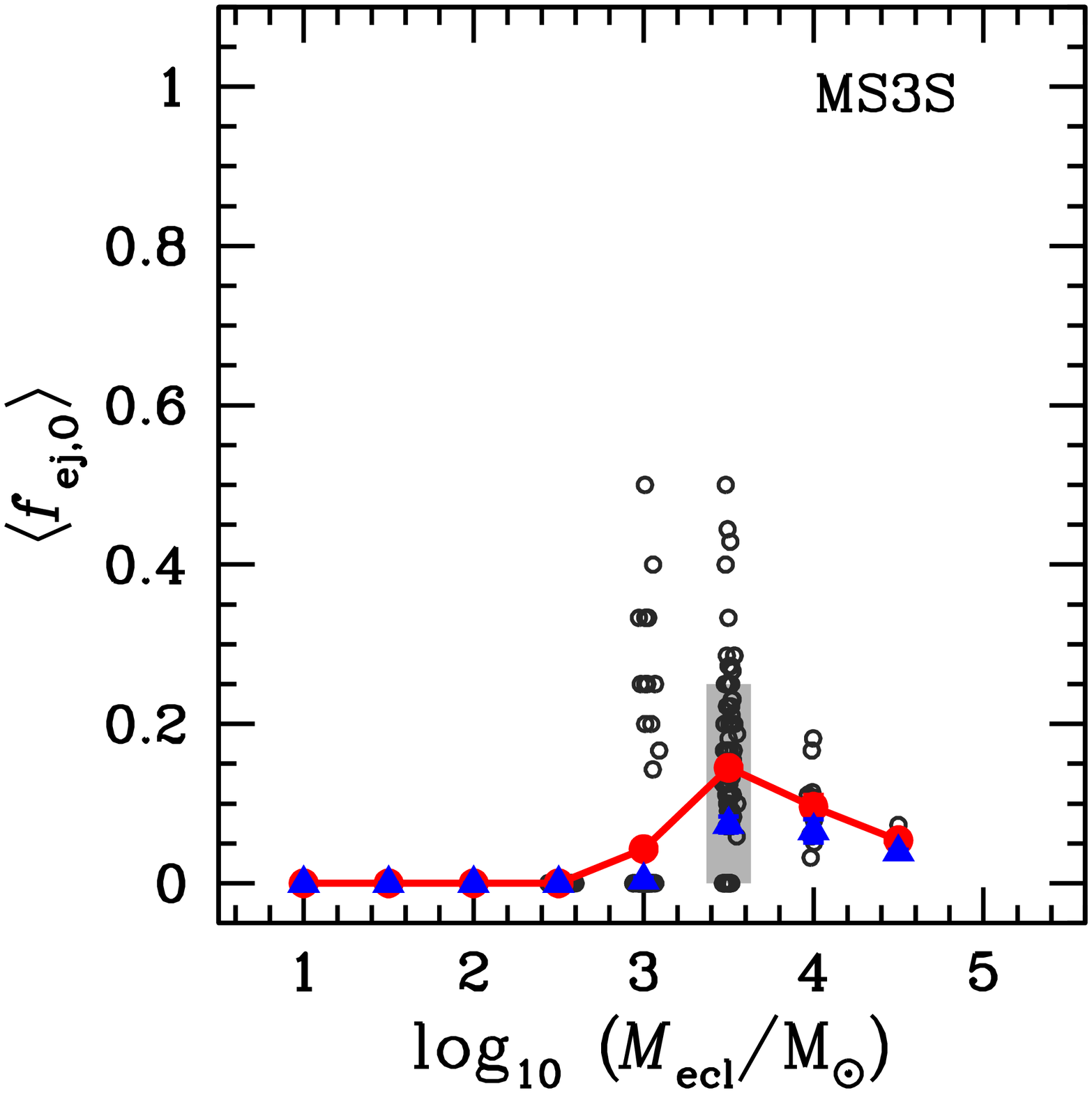}
 \includegraphics[width=59mm]{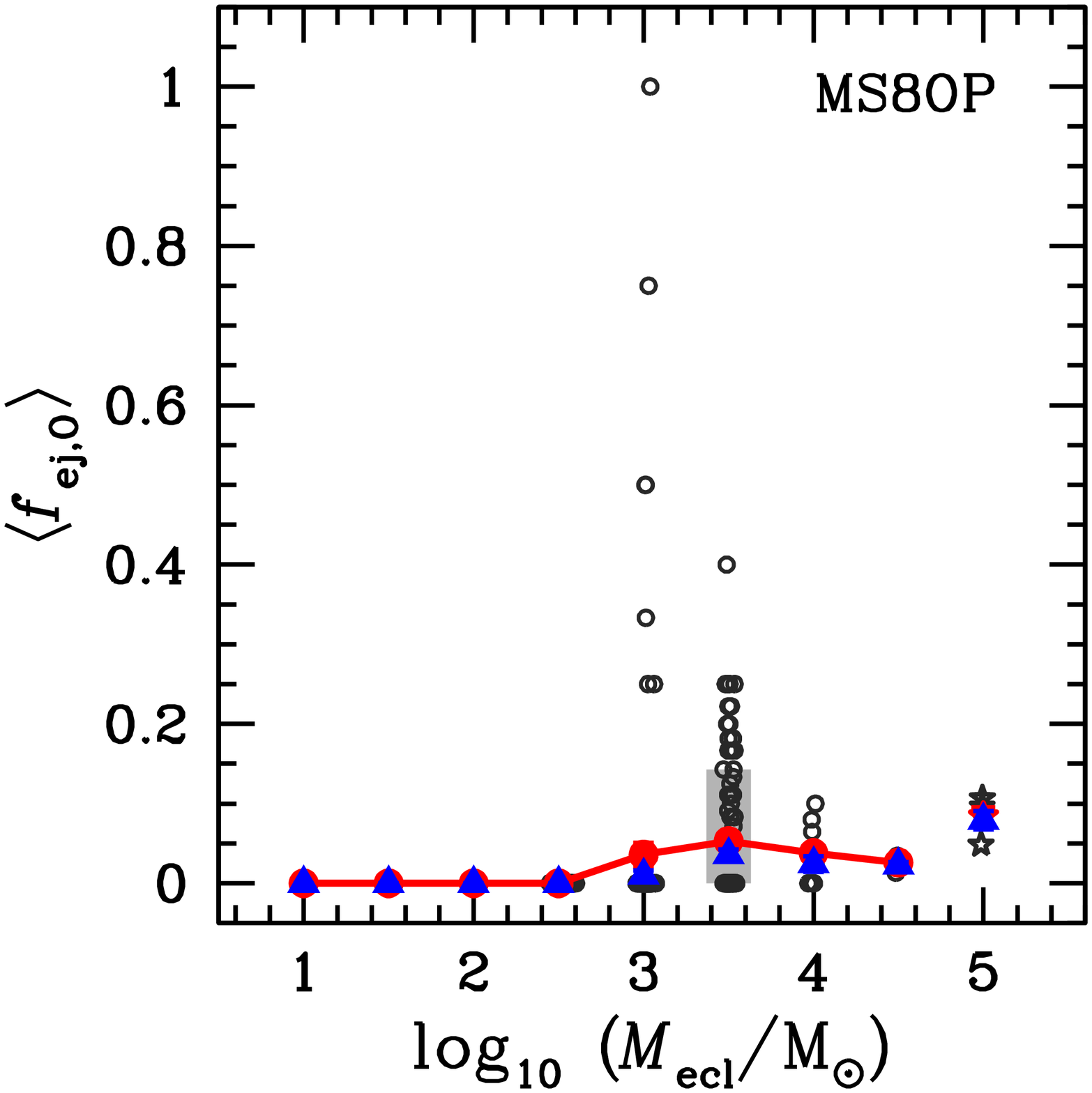} \\
 \includegraphics[width=59mm]{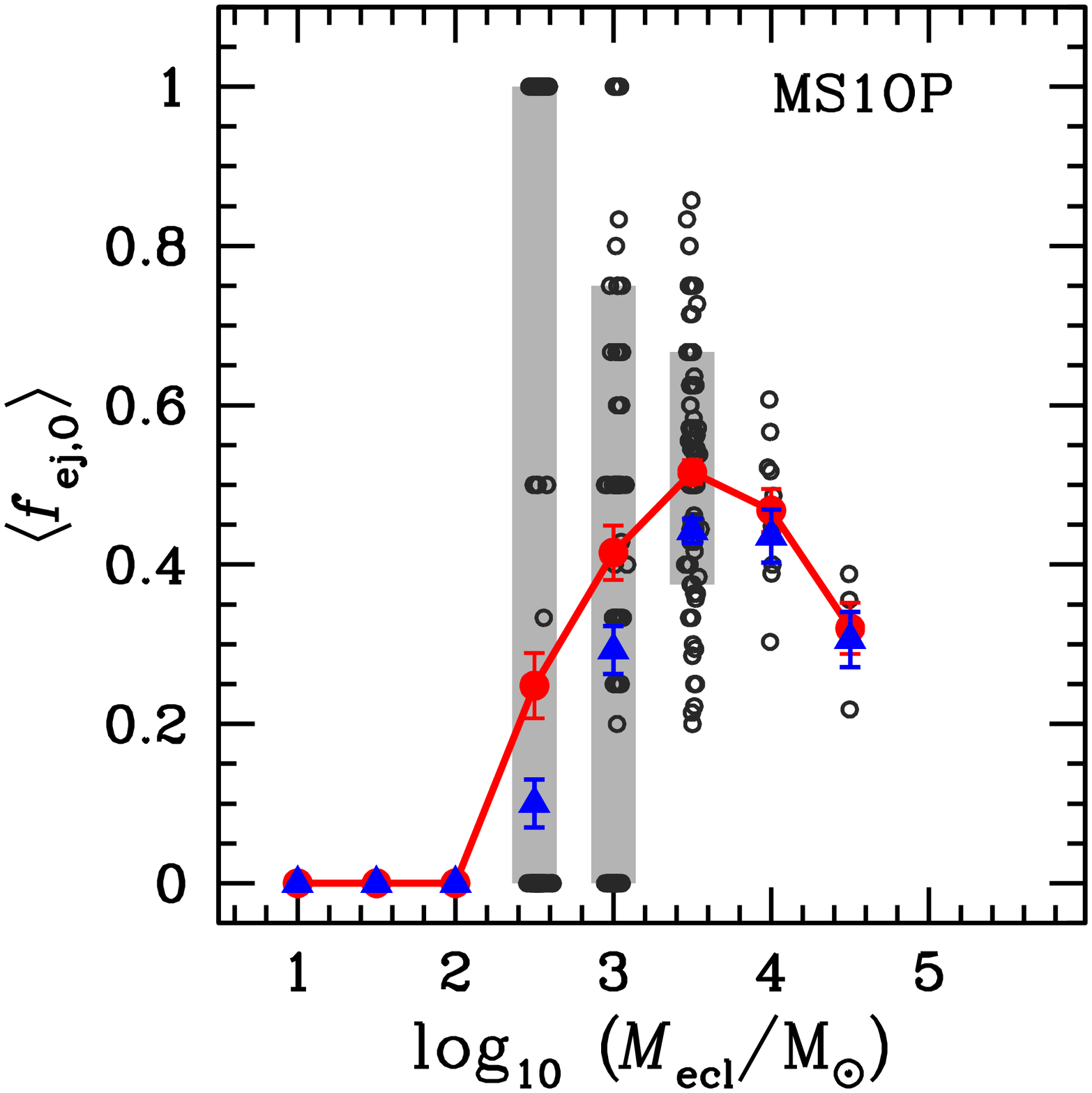}
\caption{
  Ejection fraction of O-star systems as a function of cluster mass at 3~Myr.
  Red big circles are the average O star ejection fraction for each cluster mass and
  open circles are the values of individual clusters. The (red) solid lines are drawn by connecting
  the red points to guide the dependency of the ejection fraction on the cluster mass. 
    Grey vertical bars  indicate where the central 68 per cent of the data points lie 
  (i.e., the points between 16 (lower) and 84 (upper) percentiles) 
  for $10^{3}$ and $10^{3.5}~\msun$ cluster models 
  (for the MS1OP sequence $10^{2.5}~\msun$ clusters are included). Because more than 85 per cent 
  of clusters with $\mcl=10^{2.5}~\msun$ in each sequence except for clusters in the MS1OP sequence 
  do not eject any O star, 
  the grey bar is not plotted for this cluster mass with the exception for the MS1OP sequence. 
  For more than 50 per cent of the $10^3~\msun$ clusters of all sequences, except for the MS1OP sequence
  for which only 30 per cent of the cases, the O star ejection fraction is 0.  
  Due to their small number of realisations and small spread of the ejection fraction, 
  the grey bar is not necessary for clusters with $\mcl \ge 10^{4}~\msun$.  
  Blue triangles are the average O star ejection fraction using 10 pc for the ejection criterion without
  applying a velocity criterion.
  The data for the $10^{5}~\msun$ cluster model taken from the calculations by \citet{BKO12} are 
  marked with open stars in the MS8OP model sequence panel.
  }
\label{figfej}
\end{figure*}

Throughout this paper, an O star refers to a star with a mass of $\geq 17.5~\msun$ at 3~Myr and 
an O-star system is either a single O star or a binary system  having at least one O star component.  
{\it We regard a system as being ejected when its distance from the cluster centre is greater 
than $3\times\rh$ of the cluster at 3~Myr, and its velocity is larger 
than the escape velocity at that radius.} 
Such a short distance criterion for ejections is an appropriate physical condition to quantify the real ejected 
O star fraction, in order to account for the O stars which have been dynamically ejected 
but have not travelled far from the cluster centre by 3~Myr. 
 O stars can be ejected with a velocity of a few $\kms$ from the lower-mass clusters, 
e.g., the central escape velocity of a cluster with $\mcl=1000~\msun$ and $\rh=0.3$~pc is $\approx6~\kms$. 

From the computational data the ejection fraction of O star systems, $\comp \fej$, 
is defined as the number ratio of the ejected O star systems, $\comp \nej$, 
to all O star systems that are still present in a calculation, 
$\nos$, i.e.,
\begin{displaymath}
 \comp \fej = \frac{\comp \nej}{\comp\nos}.
\end{displaymath}
Only clusters with $\mcl \geq 10^{2.5}~\msun$ are shown in the tables because clusters with 
$\mcl \leq 100~\msun$ do not have an O star in our models. Since values of $\comp\fej$ from 
individual clusters are diverse, especially for low-mass clusters, we mainly deal with 
the averaged value of $\comp\fej$, $\afej$, for the same cluster mass in a certain initial conditions set.

The average ejection fractions of O star systems for all models are shown in Figure~\ref{figfej}. 
Table~\ref{tabresult} lists averaged properties of the models, such as half-mass radius, number 
of O star systems, number of ejected O star systems, O star ejection fraction, and O star binary fraction 
in the clusters and among the ejected O star systems, at 3~Myr.  
All models show a trend that $\afej$ increases with a cluster mass up to $10^{3.5}~\msun$ 
and then decreases for a higher cluster mass. 
The average values of the ejection fraction with the same cluster mass vary from model to model. 
The peak fractions are $\approx25$ per cent for the binary-rich cluster models with $\rh=0.3$~pc 
with a weak dependency on initial mass segregation or on the initial period distribution of massive binaries, 
although the $\afej$ values of the other cluster masses change with the initial conditions. 
{\it Note that individual clusters in the mass range $10^{2.5}$ to $10^{3.5}~\msun$ may 
eject all their O star systems!}  

The ejection fractions from the initially not mass-segregated models (NMS3OP) exhibit similar 
results to the MS3OP models. 
The difference between the two models is only seen at the lowest cluster mass, $10^{2.5}~\msun$.
This can be understood because dynamical mass-segregation is inefficient for the lowest mass clusters 
due to their low density. {\it  Initial mass segregation therefore 
does not have a significant impact on massive-star ejections.}

In the models using the \citet{Set12} period distribution (MS3OP\_SP and MS3UQ\_SP), 
the same trend of the O star ejection fraction is present as above, but the $\afej$ 
values show a much broader peak than those in the model using the \citet{PK95b} 
period distribution (MS3OP). 
The $\anej$ values of the three models are similar (see Table~\ref{tabresult}). 
However, for massive clusters ($\mcl > 10^{3.5}~\msun$), MS3OP\_SP 
clusters not only have a smaller number of O star systems near 3~Myr 
due to a higher fraction of binaries and mergers, but they also eject more O star systems. 
The latter suggests that short period massive binaries are dynamically active and boost 
the ejection of O stars, especially for massive clusters. 

The numbers of O-star systems are slightly higher in the MS3UQ\_SP sequence than 
in the MS3OP\_SP sequence (see Table~\ref{tabresult}) because O-star binaries in the MS3OP\_SP models
mostly have an O-star companion while companions of O -star binaries in the MS3UQ\_SP models 
are chosen from a  broader range of stellar masses. 
For example, when the binary fraction is unity and both sequences have the same number 
of individual O stars, for the MS3OP\_SP sequence the number of O-star systems would 
be almost half of the number of individual O stars while it would larger than half 
in the case of the MS3UQ\_SP sequence. 
However, the difference in the mass-ratio distributions of the two model sequences 
does not seem to affect the trend  of the O star ejection fraction as a function of cluster mass 
having a peak at $\mcl=10^{3.5}~\msun$. The difference between the results of the 
MS3OP\_SP and MS3UQ\_SP models is only marginal. {\it Therefore, massive star 
ejections are not sensitive to the form of the mass-ratio distribution as long as 
they are paired with other massive stars, i.e., $q>0.1$.}    
 
For single-star clusters (MS3S), we find the same cluster mass at which the peak of the O star ejection fraction occurs 
although the ejection fractions are about half those of the binary-rich model sequences.
The peak appears at the same cluster mass in the sequences with different initial cluster sizes.
In the case of MS8OP ($\rh=0.8$~pc), the ejection fraction is much smaller 
than in other model sequences with $\rh=0.3$~pc. 
But $\afej$ reaches $\approx 50$ per cent in the MS1OP model sequence (initial $\rh=0.1$~pc) at the cluster mass of the peak. 
{\it The dependency of the ejection fraction on cluster mass is thus independent 
of the initial cluster radius, but more compact cluster do have higher ejection fraction of massive stars.} 

\begin{figure}
 \center
 \includegraphics[width=80mm]{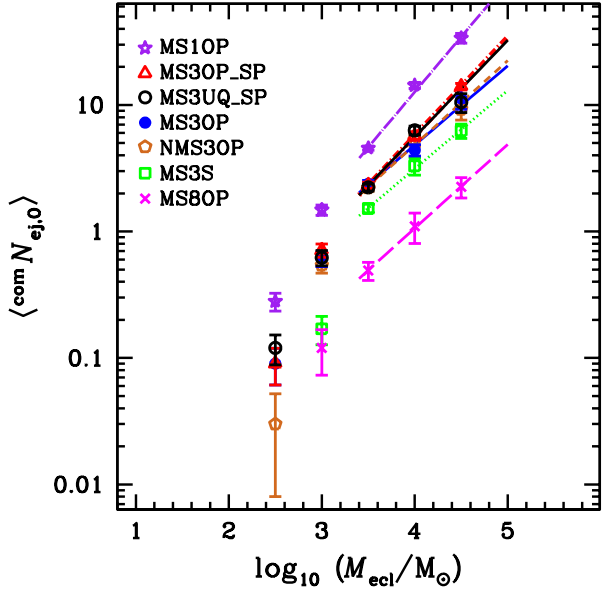}
 \caption{
 The average number of ejected O-star systems, $\anej$, as a function of cluster mass. 
 Lines are linear fits to $\anej$ with cluster mass between $10^{3.5}$ and $10^{4.5}~\msun$ (Equation~(\ref{fitnej})).
 }
 \label{Nejfig}
\end{figure}

The peak near $10^{3.5}~\msun$ of the ejection fraction can 
be understood as a consequence of the growth of the number of O stars with cluster mass 
surpassing the number of ejected O stars as the cluster mass increases and the potential smoothens. 
Figure~\ref{Nejfig} shows the average number of ejected O stars at each cluster mass. 

This implies that a-few-Myr old clusters with a mass at which the O-star ejection 
fraction peaks would present O star populations most deviating from the number of O stars 
given by the canonical IMF.  
For example the estimated stellar mass in the inner 2~pc of the  Orion Nebula Cluster (ONC) is $\approx 1800~\msun$ \citep{HH98} 
 and probably $\approx 2700~\msun$ taking account of a probable binary fraction of 50~per cent \citep{KP00}.  
 With these cluster masses, about 6--10 stars are expected to have a mass larger than 
$18~\msun$ from the canonical IMF. However, only three O-type (or $m\ge 17.5~\msun$) stars 
are found within $\approx2$~pc of the Trapezium stars \citep{LH97}. 
This discrepancy may be due to the stochastic sampling of the IMF. 
But the probability for this is less than 5 per cent when stellar masses are randomly 
drawn from the IMF with a stellar upper mass limit of $150~\msun$ (but about 10 per cent with $\mmax=50~\msun$)
 such that it is more likely that the ONC has ejected a significant fraction of its O star content. 
Furthermore, the initial mass of the cluster may have been higher than observed at the present day as 
the cluster could have lost a significant fraction of its members during the residual 
gas expulsion phase \citep{KAH01}. Thus the ONC formed with a mass close to the peak 
mass we found in this study and could have lost a significant fraction of its massive stars 
through the dynamical ejection process. This may explain why the ONC's massive star population 
deviates from the canonical IMF \citep{PK06}.  Examples of such events would be 
two O-type single runaway stars, AE Aur and $\mu$ Col, and a massive binary system $\iota$~Ori 
which may have been ejected from the ONC through a binary-binary interaction \citep{GB86,Hdd01,GPE04}. 
Future proper motion surveys, such as with the \facility{Gaia} mission, may provide further constraints 
on this question, because the putative ejected O stars from the ONC would have to become evident.

Very massive clusters may have small ejection fractions of O stars but 
their very upper stellar mass end could lack stars because the ejection fraction 
of stars more massive than $100~\msun$ can be from $\approx 10$ to $\approx100$ 
per cent \citep{BKO12}.  Furthermore, they can eject not only very massive single stars but also 
very massive binaries \citep{OKB14}.  This bias through the most massive stars being 
ejected from star-burst clusters and the observed IMF of star-burst clusters being close 
to the canonical value \citep[e.g., the R136 cluster in the Large Magellanic Cloud,][]{MH98} 
may imply the IMF to be top-heavy in star-burst clusters \citep{BK12}. There is some evidence for a possibly systematically varying IMF above a few $\msun$:  A number of independent arguments have shown the IMF to become increasingly top-heavy with increasing star-formation rate density on a pc scale \citep{Markset12}, with a dependency on the metallicity. 
This metallicity dependence is such that metal-rich environments tend to produce a steeper (i.e. top-light) IMF slope. Evidence for this may have emerged in M31 data \citep{Weisz15}. 
The models calculated here are, however, in the invariant regime. 

\begin{figure}
 \center
 \includegraphics[width=70mm]{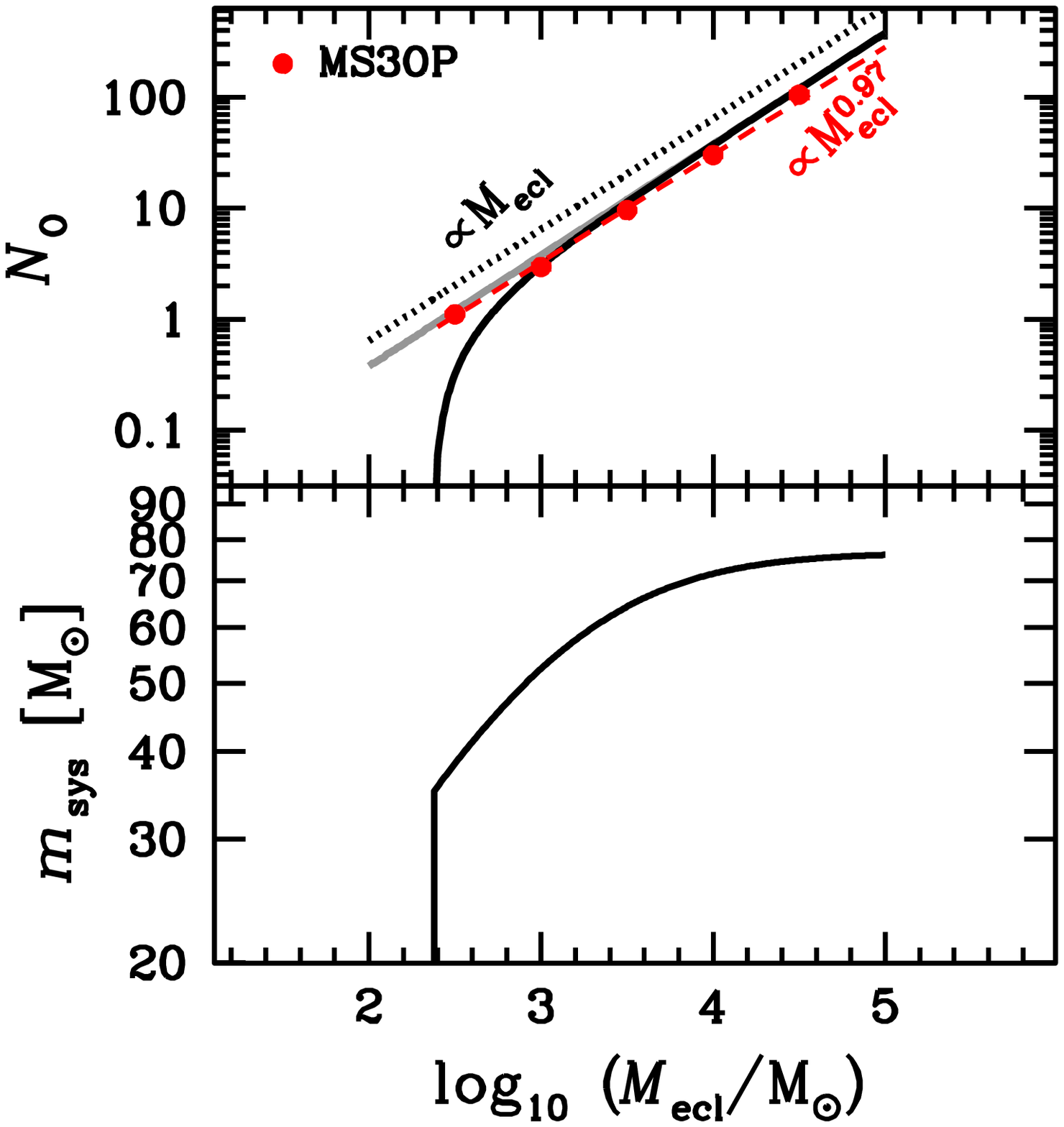}
 \caption{Top: Number of O stars in a cluster calculated using the IMF, $N_{\mathrm{O,IMF}}$ 
 (Equation~(\ref{eqno}), solid line). 
 The dotted line indicates $\nos \propto \mcl$. Red filled circles present the average number of
 O-star systems from \nbody\  calculations, $\anos$, in the MS3OP model sequence. 
 A red dashed line indicates $\anos \propto \mcl^{0.97}$ obtained by a linear fit of 
 $\log_{10} \anos$ -- $\log_{10}\mcl$ to the MS3OP model sequence. 
 The other two model sequences, NMS3OP and MS3S, have almost the same slopes as the MS3OP model sequence (Equation~(\ref{eqnos})).   The grey solid line is the number of O stars derived from the IMF with the universal upper mass limit of $150~\msun$.
 Bottom: Assumed system mass of an O star binary, $\msys$.  
 The values are twice the average O star mass in a cluster calculated from the IMF (Equation~(\ref{imf})). 
 Here we assumed that an O star binary is an equal mass binary. 
 Although the mass-ratio distribution of O star binaries is close to a uniform distribution for $q>0.1$ 
 in observations \citep{Set12}, our assumption is sufficient to show the dependence 
 of the average system mass of O star binaries on cluster mass.   
 }
 \label{Nofig}
\end{figure}

A theoretically expected relation between the ejection fraction and the cluster mass can be found. 
First, the number of O stars in a cluster, $N_{\mathrm{O,IMF}}$, which is a function of 
$\mmax(\mcl)$ and $\mcl$, is
\begin{eqnarray}
N_{\mathrm{O,IMF}}&=& \int^{\mmax(\mcl)}_{17.5~\msun} \xi(m) \dm, \nonumber \\
&\approx &\frac{3.390\times 10^{-4}-0.014\mmax^{-1.3}}{0.100-0.062\mmax^{-0.3}}\mcl,
\label{eqno}
\end{eqnarray}
by using the IMF (Equation~(\ref{imf})).  
The normalisation constant, $k$, in Equation~(\ref{imf}) is calculated from $\mcl=\int^{\mmax}_{0.08~\msun}m\xi(m)\dm$. 
The top panel of Figure~\ref{Nofig} presents $N_{\mathrm{O,IMF}}$ as a function of $\mcl$ from Equation~(\ref{eqno}).
Since we adopt the $\mmax$--$\mcl$ relation, Equation~(\ref{eqno}) is not a simple linear function 
of cluster mass especially for low-mass clusters in which $\mmax$ is highly dependent on the cluster mass.
$N_{\mathrm{O,IMF}}$, however, becomes almost linearly proportional to $\mcl$ 
for massive clusters where $\mmax$ is saturated (see the upper panel of Figure~\ref{Nofig}).
Note that the solid line in the upper panel of Figure~\ref{Nofig} and the numbers in 
Table~\ref{tabresult} are slightly different because Equation~(\ref{eqno}) 
gives the number of all {\it individual} O stars from the IMF,   
whereas $\anos$ listed in Table~\ref{tabresult} is the number of O star {\it systems} at 3~Myr. 
The difference between the number of O stars from the IMF, $N_{\mathrm{O,IMF}}$,
and from our \nbody\ calculations, $\langle \comp \nos \rangle$, at the cluster mass of 
$10^{2.5}~\msun$ in Figure~\ref{Nofig} arises due to our choice of forcing a cluster to have at least
one $\mmax$-star initially. This would not have been required if optimal sampling \citep{Ket13} 
had been available when the theoretical young star cluster library was computed pre-2013.  

Our stellar masses are randomly drawn from the IMF with the upper mass limit given by the $\mmax$--$\mcl$ relation. 
Therefore, our sampling would produce a slightly lower maximum stellar mass in a cluster but a similar number of O stars compared to what random sampling without the $\mmax$--$\mcl$ relation would give. 
The grey line in the upper panel of Figure~\ref{Nofig} is the expected number of O stars 
from the IMF with a universal upper stellar-mass-limit of $150~\msun$ for all cluster masses. 
The figure indicates that a higher number of O stars is expected in the IMF with a universal 
and constant $\mmax=150~\msun$ than with the $\mmax$--$\mcl$ relation at lower cluster mass 
($\mcl \lesssim$ a few $1000~\msun$, the black solid line in the same figure). 
But the values of $\anos$ in our \nbody\ models are close to the grey line because of adding 
one $\mmax(\mcl)$-star in a cluster of mass $\mcl$ by default. 
So whether adopting the relation or not would have little effect on the number of O stars in our models. 
The proper comparison for outcomes of the two assumptions ($\mmax=150~\msun$ or the $\mmax$--$\mcl$ relation) 
can be done by performing a large set of Monte Carlo calculations. However, it is beyond the objectives of this study.   In any case, the $\mmax$--$\mcl$ relation is observationally well supported 
for young star clusters and star-forming regions in the Milky Way given the most recent data \citep{KM11,WKP13}. 

Fitting the number of O-star systems present, regardless of whether the systems are in the star cluster or ejected, in \nbody\ calculations at 3~Myr to a simple power law gives,
\begin{equation}
\anos \propto \left\{ 
 \begin{array}{ll}
 \mcl^{0.97\pm0.01},& \mathrm{MS3OP},\\
 \mcl^{0.96\pm0.01},& \mathrm{NMS3OP},\\
 \mcl^{0.96\pm0.01},& \mathrm{MS3OP\_SP},\\
 \mcl^{0.96\pm0.01},& \mathrm{MS3UQ\_SP},\\
 \mcl^{1.01\pm0.01},& \mathrm{MS3S},\\
 \mcl^{0.95\pm0.01},& \mathrm{MS8OP},\\
 \mcl^{1.00\pm0.01},& \mathrm{MS1OP}.
 \end{array} \right.
\label{eqnos}
\end{equation}
The numbers of O star systems in all model sequences are more or less linearly proportional to the cluster mass 
as expected from the IMF despite various binary fractions. 

Next, the relation between the number of ejected O stars and cluster mass is needed. 
According to \citet{Ve03}, the binary -- single star scattering encounter rate, $\Gamma$, is
\begin{equation}
\Gamma \propto \frac{\rho_{0}^{2}r_{\mathrm{c}}^{3}}{\vcl}a,
\label{eqga}
\end{equation}
where $\rho_{0}$ is the central mass density, $\vcl$ the velocity dispersion of the cluster, 
$r_{\mathrm{c}}$ the core radius, and $a$ the semi-major axis of the binary. 
The semi-major axis can be replaced with the system mass, $\msys$,
and the orbital velocity, $v_{\mathrm{orb}}$, of the binary by using Equation (8.142) in 
\citet{Kr08} as,
\begin{equation}
a \propto \msys v_{\mathrm{orb}}^{-2}.
 \label{eqa}
\end{equation}
Considering binaries with a circular velocity similar to the velocity dispersion of the cluster are
dynamically the most active (i.e., $v_{\mathrm{orb}} = \vcl$) and $\vcl\propto \sqrt{\rho_{\mathrm{0}}}r_{\mathrm{c}}$, 
Equation~(\ref{eqga}) becomes
\begin{equation}
\Gamma \propto \rho_{0}^{0.5} \msys \propto \mcl^{0.5} \msys ,
\label{eqga2}
\end{equation}
since all our cluster models in a given sequence have the same initial half-mass radius 
the initial central density is only proportional to the cluster mass. 
A binary ejecting an O star after a strong interaction with a single star 
 is likely a binary consisting of O stars since the least massive star is generally 
ejected after a close encounter between a binary and a single star.
Therefore, $\msys$ can be approximated to twice the average mass of O stars in a cluster.
We assume, 
\begin{eqnarray}
\msys&=& 2\langle m_{\ge17.5\msun}\rangle = 2\frac{\int^{\mmax}_{17.5}m\xi(m)\dm}
 {\int^{\mmax}_{17.5}\xi(m)\dm} \nonumber\\  
& \approx & 8.66\frac{\mmax^{-0.3}-0.42}{\mmax^{-1.3}-0.02},
\label{eqmsys}
\end{eqnarray}
using the IMF (Equation~(\ref{imf})).
Thus $\msys$ is a function of $\mmax$ so that $\msys =\msys(\mmax)$. 
Due to the $\mmax$--$\mcl$ relation, $\msys$ is thus a function of the cluster mass 
(the lower panel of Figure~\ref{Nofig}). 

The number of ejections is proportional to the encounter rate,
\begin{equation}
\suth \nej \propto \Gamma \propto \msys(\mmax)\mcl^{0.5}.
\label{eqNej}
\end{equation}
The $\mmax$--$\mcl$ relation has only a minor effect on the number of ejected O stars 
for the massive clusters, i.e., the dependence of $\msys$ on $\mcl$ becomes negligible in Equation~(\ref{eqNej}).  
$\suth \nej$ is thus expected to be proportional to $\mcl^{0.5}$ for the massive clusters. 
We fit $\anej$ as a power-law of cluster mass for clusters with 
$10^{3.5} \le \mcl/\msun \le 10^{4.5}$ (Figure~\ref{Nejfig}). 
These fits provide for each model sequence,
\begin{equation}
 \anej \propto \left\{
 \begin{array}{ll}
 \mcl^{0.63\pm0.066},& \mathrm{MS3OP},\\
 \mcl^{0.67\pm0.066},& \mathrm{NMS3OP},\\
 \mcl^{0.78\pm0.034},& \mathrm{MS3OP\_SP},\\
 \mcl^{0.76\pm0.061},& \mathrm{MS3UQ\_SP},\\
 \mcl^{0.62\pm0.061},& \mathrm{MS3S},\\
 \mcl^{0.66\pm0.100},& \mathrm{MS8OP},\\
 \mcl^{0.87\pm0.100},& \mathrm{MS1OP}.
 \end{array} \right.
 \label{fitnej}
\end{equation}
All model sequences but MS3OP\_SP and MS1OP exhibit slopes similar to each other within an error, 
in a range of 0.62--0.67. 
The power-law exponents we obtain from our \nbody\ models are 
slightly steeper than 0.5 in Equation~(\ref{eqNej}) but with only a marginal difference. 
It can be argued that this is due to a broad range of binary parameters for which binaries could 
have engaged in the ejection of O stars in our models and the complexity of the few-body scattering process. 
The model sequence MS3OP\_SP shows a steeper slope, 
i.e., the more massive clusters in this model eject O stars more 
efficiently compared to other models. As the model sequence MS1OP begins 
with an extreme density the results of the model
significantly deviate from the simplified theoretical expectation with simple approximations. 

\begin{figure}
 \center
 \includegraphics[width=80mm]{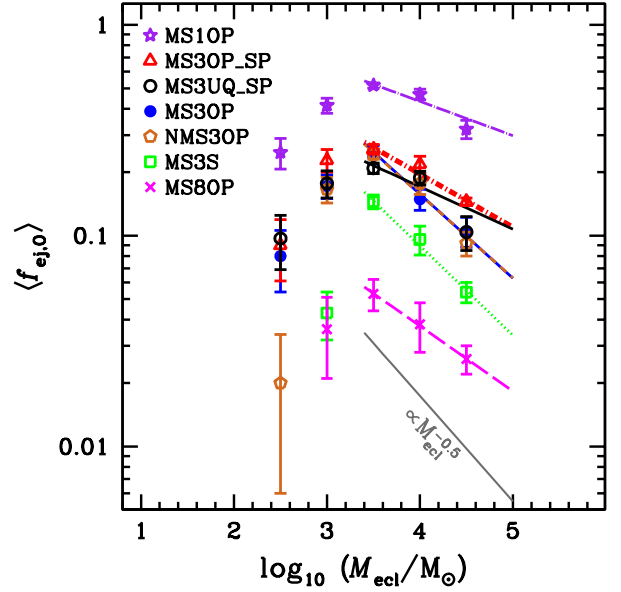}
 \caption{
 The average ejection fraction of O-star systems.  
 Lines are linear fits of the average ejection fraction from the clusters with 
  $\mcl=10^{3.5}$--$10^{4.5}~\msun$ (Equation~(\ref{fitfej})). 
 The grey solid line shows the relation $\suth \fej \propto \mcl^{-0.5}$.
 }
 \label{fefit}
\end{figure}

Finally, the theoretically expected ejection fraction of O stars is
\begin{equation}
\suth  \fej = \frac{\suth \nej}{N_{\mathrm{O,IMF}}} \propto F(\mmax)\,\mcl^{-0.5},
\label{eqfej}
\end{equation}
where $F({\mmax}$) comes from the part of Equations~(\ref{eqno}) and (\ref{eqNej}) in which $\mmax$ plays a role.  
For massive clusters ($\mcl\ge 10^{3.5}~\msun$), $F(\mmax)$ in Equation~(\ref{eqfej}) becomes 
very weakly dependent on the cluster mass  as in Equation~(\ref{eqNej}), and thus this term can be ignored. 
Therefore, $\suth \fej$ is expected to be proportional to $\mcl^{-0.5}$ in this cluster mass regime.
Figure~\ref{fefit} shows the linear fits to the O star ejection fractions as a function
of cluster mass for the massive clusters ($10^{3.5}\le \mcl/\msun \le 10^{4.5}$) in a log-log scale. 
Within a cluster mass range $10^{3.5}$--$10^{4.5}~\msun$, the numerical data suggest the relation between 
O star ejection fraction and cluster mass as follows,
\begin{equation}
\afej \propto \left\{
 \begin{array}{ll}
 \mcl^{-0.40\pm0.064},& \mathrm{MS3OP},\\
 \mcl^{-0.40\pm0.055},& \mathrm{NMS3OP},\\
 \mcl^{-0.25\pm0.029},& \mathrm{MS3OP\_SP},\\
 \mcl^{-0.20\pm0.060},& \mathrm{MS3UQ\_SP},\\
 \mcl^{-0.43\pm0.055},& \mathrm{MS3S},\\
 \mcl^{-0.31\pm0.094},& \mathrm{MS8OP},\\
 \mcl^{-0.16\pm0.036},& \mathrm{MS1OP}.
 \end{array} \right.
 \label{fitfej}
\end{equation}
The results of most models are reasonably consistent with the relation between the ejection fraction 
and cluster mass expected from the binary-single star scattering encounter rate (Equation~(\ref{eqfej})) 
when applied to massive clusters in which the maximum stellar mass is less sensitive to cluster mass. 

For lower mass clusters, the number of O stars in a cluster and the ejection rate do not 
simply follow from the above equations due to the dependency of $\mmax$ on 
the cluster mass and the highly collisional nature of the cluster cores which contain but a few  O stars.  

The ejection fractions from individual clusters are plotted with open circles in Figure~\ref{figfej}.
For lower mass clusters the O star ejection fraction varies largely from one cluster to another, 
e.g., the values spread from 0 to 1 for clusters with $\mcl \leq 10^{3.5}~\msun$. 
The spread gets smaller with increasing cluster mass as the stochastic effect diminishes with 
a larger number of O stars in more massive clusters. For massive clusters ($\mcl\geq 10^{4}~\msun$) 
the individual ejection fractions mostly lie close to the average value. 

Only for a comparison, we additionally show results using 10~pc as the ejection criterion in Figure~\ref{figfej}
(without a constraint on the velocity of the stars) since a larger distance criterion is more appropriate from an observational point of view. 
The ejection fractions using 10~pc as a distance criterion show a similar shape and the peak 
at the same cluster mass, only the values are smaller (max. up to $\approx 17$~per cent for clusters with an initial $\rh=0.3$~pc) 
than the results using our main ejection criteria. The difference becomes smaller or almost disappears for the massive clusters.
Massive clusters may eject O stars most efficiently at earlier time of the evolution due to their short dynamical (i.e., crossing) time-scale, and soon their ejection efficiency would drop 
as the O star population is reduced in the core of the cluster. 
Also, in the massive clusters the velocity of the ejected systems is generally high (see Section~5.1) so they quickly move further than 10~pc away 
from the cluster centre. Low-mass clusters have longer dynamical times-scales and they eject O stars moving 
relatively slowly so that they need more time to reach a distance of 10~pc.

\section{Field O stars from dynamical ejection processes}
\begin{figure}
 \center
 \includegraphics[width=80mm]{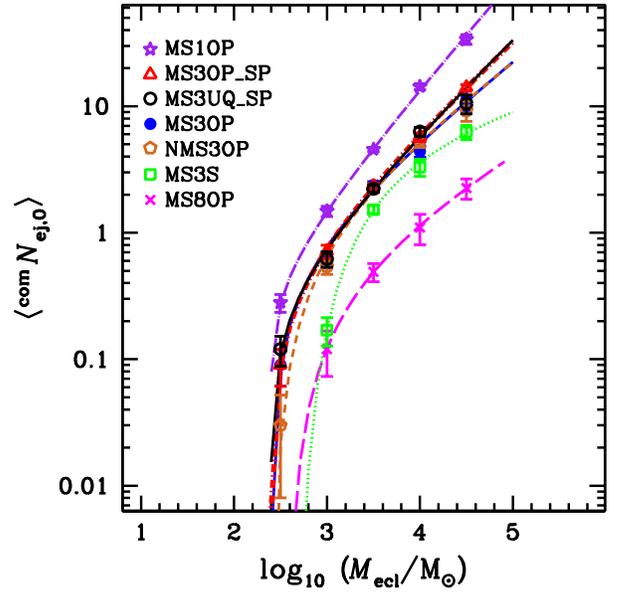}
 \caption{
 Fitting function for $\anej$ (Equation~(\ref{eqfit}) and Table~\ref{fitpar}).
 The symbols and line styles are the same as in Figure~\ref{Nejfig}.
 }
 \label{fit}
\end{figure}
 
Our \nbody\ results show that $\log_{10}\anej$ steeply increases with $\log_{10}\mcl$ at a lower cluster-mass 
range ($\mcl<10^{3.5}~\msun$) while it can be approximated to a linear function of $\log_{10}\mcl$ at a higher 
cluster-mass range. 
In the previous section, we use data only from the clusters with $10^{3.5}~\msun \le \mcl
\le 10^{4.5}~\msun$ to fit a linear relation between $\log_{10}\anej$ and $\log_{10}\mcl$ (Figure~\ref{Nejfig}).
In order to fit the number of ejected O stars for the entire cluster mass range in which 
$\anej$ is $> 0$, we use the following functional form, 
\begin{equation}
\log_{10}y = a + b\log_{10}x -\frac{1}{(\log_{10}x- \log_{10}x_{0})^{n}}, 
\label{eqfit} 
\end{equation}
where $x>x_{0}$, $y = \anej$ and $x=\mcl/\msun$.
In this function, $\log_{10}y$ converges to $-\infty$ for $x \rightarrow x_{0}$ 
and approximates a linear function, $a+b\log_{10}{x}$, for $x \gg x_{0}$.
This function can thus be fitted to our results. 
The value $x_{0}$ indicates the cluster mass where the $y$ value, i.e., $\anej$, $\approx0$. 
For all models but MS3S and MS8OP, we choose $x_{0}=240~\msun$ which is the maximum cluster mass at which clusters 
do not host single O stars initially, as given by the $\mmax$--$\mcl$ relation.
While for the other two models we choose $x_{0}=10^{2.5}~\msun$ at which mass a cluster contains O stars 
but $\anej$ is 0 in the N-body calculations.
By performing non-linear least squares fitting of Equation~(\ref{eqfit}) to our results,  
we obtain the parameters listed in Table~\ref{fitpar} and plot the results in the upper panel of Figure~\ref{fit}. 

\begin{deluxetable}{lcccc}
\tablewidth{0pt}
\tablecaption{Parameters by fitting Equation~(\ref{eqfit}) for all models.\label{fitpar} }
\tablehead{ \colhead{Model} & \colhead{$a$} & \colhead{$b$} & \colhead{$n$}} 
\startdata
MS3OP       & $-0.56\pm0.25$ & $0.53\pm0.07$ & $0.29\pm0.04$ \\ 
NMS3OP      & $-0.43\pm0.25$ & $0.49\pm0.07$ & $0.42\pm0.05$ \\ 
MS3OP\_SP   & $-0.91\pm0.16$ & $0.64\pm0.04$ & $0.26\pm0.04$ \\ 
MS3UQ\_SP   & $-1.04\pm0.25$ & $0.67\pm0.07$ & $0.22\pm0,04$ \\
MS3S        & $0.67\pm0.32 $ & $0.14\pm0.08$ & $0.90\pm0.10$ \\ 
MS8OP       & $-0.60\pm0.01$ & $0.37\pm0.003$& $0.51\pm0.03$ \\ 
MS1OP       & $-1.17\pm0.16$ & $0.80\pm0.05$ & $0.16\pm0.03$ 
\enddata
\end{deluxetable}

\begin{figure}
\center
\includegraphics[width=85mm]{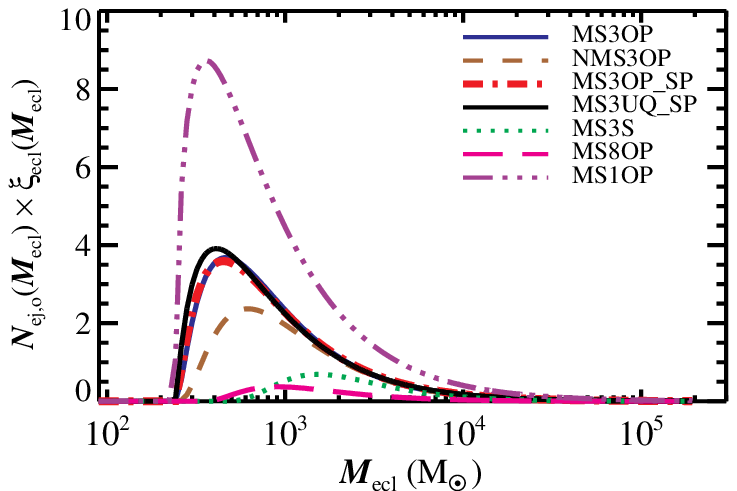}\\
\includegraphics[width=85mm]{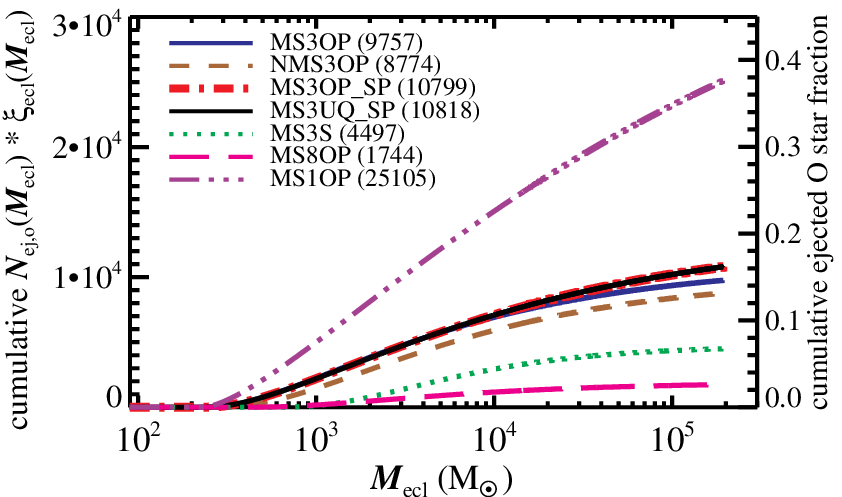}
\caption{Top: Total number of the ejected O star systems per cluster mass in the Galaxy 
 (Equation~(\ref{nofd})). For example, according to the most realistic model sequence 
 (MS3UQ\_SP), in total about 4 ejected O star systems per 10 Myr are contributed from all clusters 
 weighing about $400~\msun$. 
 The small numbers of ejected O star systems contributed from massive clusters 
 come about because the number (or probability) of such massive clusters to form is small 
 according to the very young or embedded cluster mass function (Equation~(\ref{eqcmf})). 
 Bottom: Cumulative number of the ejected O star systems as a function of cluster mass. Right y-axis 
 shows the ejected O star fraction of the total O stars formed within 10~Myr deduced by using Milky Way type parameters.
 Numbers next to the model sequence names are the expected total number of ejected O star systems which 
 are the sum of Equation~(\ref{nofd}) over all clusters for each model sequence. 
 For example, the MS1OP model sequence would imply that all clusters formed within 10~Myr eject 
 in total almost $2.5\times 10^{4}$ O star systems assuming the $\mathrm{SFR}=3~\msun~\mathrm{yr}^{-1}$ such that $\mclmax=1.9\times10^{5}~\msun$. 
  As another example, according to the most realistic sequence of models 
 (MS3UQ\_SP), every 10~Myr, 10818 O star systems are ejected from all clusters 
 formed in this time interval, clusters with a mass of almost $400~\msun$ 
 have the largest contribution to this population (upper panel) 
 and clusters with masses up to $10^{3}~\msun$ contribute almost 33 per cent to this population (lower panel, right axis), 50 per cent of all ejected O stars originate from $\mcl \le 3\times10^{3}~\msun$ clusters.}
\label{nmcl}
\end{figure}

The mass distribution of young star clusters can be described with a simple power-law, 
\begin{equation}
\xi_{\mathrm{ecl}}(\mcl)= k_{\mathrm{ecl}} \, \mcl^{-\beta},
\label{eqcmf}
\end{equation}
where the normalisation constant $k_{\mathrm{ecl}}$ is calculated from the total mass of 
all clusters which form within a certain time. 
Assuming a global Milky Way star formation rate (SFR) of $3~\msun~\mathrm{yr}^{-1}$, a minimum cluster mass 
$\mclmin=10~\msun$, and $\beta=2$ for the Milky Way, 
the maximum cluster mass, $\mclmax$, that can form in the Galaxy becomes 
$\approx 1.9\times 10^{5}~\msun$  \citep[and references therein]{WKL04,MK11}. 
The total mass formed in stars over time $\delta t$, $M_\mathrm{SCS}$, is 
\begin{equation}
M_{\mathrm{SCS}} = \mathrm{SFR} \times \delta t = \int^{\mclmax}_{\mclmin} 
       \mcl \, \xi_{\mathrm{ecl}}(\mcl) \, d\mcl
\end{equation}
where $\delta t$ is the cluster-population formation time-scale, about 10~Myr \citep{WKL04}.
 We adopt here the $\mclmax$--SFR relation derived by \citet{WKL04}, i.e., the mass 
of the most massive forming cluster is dependent on the physical properties of the star-formation environment provided 
by a self-regulated galaxy. It has been found that the luminosity of the brightest cluster increases 
with the galaxy-wide SFR \citep[and references therein]{LS02} or with the total number of clusters \citep{WB03}. 
The relation is treated to be the result of the size of the sample in several studies 
(\citealt{WB03}; see also section 2.4 of \citealt{PMG10}, and references therein). 
Recently, \citet{PGK13} ruled out that the relation results from the pure size-of-sample 
effect by studying a radial dependancy of maximum star cluster masses in M33.  
 \citet{Ret13} confirm this by studying the $\mclmax$ versus SFR relation in their independent observational survey. 
The authors noted that their result can be explained with purely random sampling but 
only if the luminosity function of super star clusters at the bright end is very steep 
(a slope $> 2.5$), steeper than usually observed ($\approx2$). 

By combining the fitting function for the number of ejected O stars (Equation~(\ref{eqfit})) 
with the cluster mass function (Equation~(\ref{eqcmf})), 
the number of O stars being ejected to the field contributed by clusters in the mass range $\mcl$ to $\mcl+d\mcl$ is
\begin{equation}
 d\tot \nej (\mcl) = \anej(\mcl) \times \xi_{\mathrm{ecl}}(\mcl) \, d\mcl, 
\label{nofd}
\end{equation}
and is shown in the upper panel of Figure~\ref{nmcl}.  
As the number of clusters decreases more steeply ($\xi_{\mathrm{ecl}}(\mcl) \propto \mcl^{-2}$) 
than the number of ejected O star systems from a cluster increases with cluster mass ($\anej(\mcl)\propto\mcl^{\gamma}$ with $0.6<\gamma<0.9$, Section~3),
the number of O stars contributed to the field by dynamical ejections is a function 
decreasing with increasing cluster mass.  The most prominent contributors to 
the population of ejected O stars are $\approx$400$~\msun$ clusters.

In the lower panel of Figure~\ref{nmcl}, cumulative numbers of ejected O-type systems as a function of cluster mass are shown.
The fraction (right Y-axis) is calculated by dividing the cumulative number by the total number of O stars formed during 
the cluster-population forming time scale of 10~Myr, which can be deduced from Equations (\ref{eqno}) and (\ref{eqcmf}),
\begin{equation}
\tot N_{\mathrm{O}} = \int^{M_{\mathrm{ecl,max}}}_{M_{\mathrm{ecl,min}}} 
 N_{\mathrm{O,IMF}}(\mcl) \, \xi_{\mathrm{ecl}}(\mcl) \, d\mcl.
\label{eqtno}
\end{equation}
The parameters which we adopt here for the Milky Way provide $\tot N_{\mathrm{O}}\approx 66749$ being formed every 10~Myr. 
Note $\tot N_{\mathrm{O}}$ is the total number of {\it individual} O stars which have formed within 
the formation time scale, while $d\tot \nej(\mcl)$ is the number of ejected O star {\it systems}.
It is difficult to derive the total number of O star systems from the models because a binary fraction evolves due to 
dynamical disruptions/captures and mergers between stars, the final binary fraction varies 
with cluster mass and initial conditions, and some O stars have an O star secondary while some do not.
The fraction shown in the lower panel of Figure~\ref{nmcl} can, thus, be interpreted as being a minimum value since 
the total number of O star systems would be smaller than $\tot N_{\mathrm{O}}$ which counts all individual 
O stars from the IMF and cluster mass function.    
Our models thus yield as a lower boundary that binary-rich clusters with an initial $\rh=0.3$~pc can eject 
$\approx$13--16 per cent of the total O star systems and even $\approx38$ per cent 
can be dynamically ejected if clusters form more compact, 
e.g., with an initial $\rh=0.1$~pc (Figure~\ref{nmcl}). 
About 50 per cent of ejected O stars come from clusters with $\mcl\lesssim 3000$--$5000~\msun$.
 
The fraction of O stars found in clusters and/or OB-associations is 70--76 per cent 
\citep[][depending on how to count a binary]{Wet05,ELT11} 
yielding the fraction of the field O stars as being 24--30 per cent. 
Our model is thus reasonably consistent with these data. 
Additional three processes that are not part of the present model but which 
increase the apparent number of field O stars are : 
 (i) Some of the observed field stars may originate from clusters which contained only a few or one O stars 
   and which dissolved at an early age by rapid gas removal \citep{Get12}. That is such O stars 
 ``lost'' their birth cluster through disruptive gas removal \citep[see also][]{KB02,Weidner07}. 
 (ii) Small-$N$ clusters for which the two-body relaxation time is comparable to the crossing time 
 also dissolve within a few initial crossing times \citep{Met12}, which is aided by stellar evolution within a few tens of Myr.
 (iii) Some O stars may be ejected when they are companions to primaries that explode 
 as supernovae at an age older than 3~Myr \citep{ELT11}.

\begin{figure}
\center
\includegraphics[width=85mm]{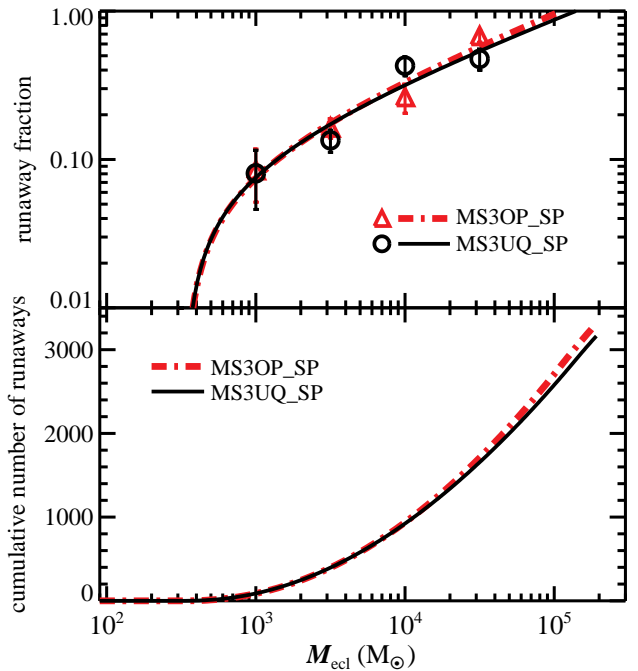}
\caption{ Top: Runaway fraction among the ejected O stars from two model sequences, MS3OP\_SP and MS3UQ\_SP. The data are fitted with the same functional form of Equation~(\ref{eqfit}) (curved lines). Error bars indicate Poisson errors.  
Bottom: Expected cumulative number of runaways as a function of cluster mass. The numbers are derived by multiplying the fitted lines in the upper panel to the number of ejected O star systems with cluster mass,
Equation~(\ref{nofd}). More than 70 per cent of O star runaways are expected to originate in the clusters more massive than $10^4~\msun$. }
\label{rafig}
\end{figure}

In observations about 6 per cent of O stars are found being runaways which have a velocity higher $30~\kms$ \citep{ELT11}. 
 The runaway fraction among the dynamically ejected O stars generally increases 
with cluster mass (upper panel of Figure~\ref{rafig}), e.g., 0 ($\mcl=10^{2.5}~\msun$) to 43 ($\mcl=10^{4.5}~\msun$) per cent in the case of the MS3UQ\_SP model 
and 81 percent for the $10^5~\msun$ model in \citet{BKO12}. We deduce the number of the runaways first 
by fitting a function for the runaway fraction as a function of cluster mass with the same functional form of the number of ejected O stars (Equation~(\ref{eqfit}), but $y$ being 
the runaway fraction) and then by multiplying the fitting function 
by the number of ejected O stars (Equation~(\ref{nofd}), 
upper panel of Figure~\ref{nmcl}). By dividing the total number of the runaways 
by the total number of the ejected stars (lower panel of Figure~\ref{nmcl}), 
we obtain the runaway fraction among the ejected O stars. Our most realistic model, MS3UQ\_SP, 
predicts about 30 per cent of the ejected O stars, 
i.e., about 5 per cent of all O star systems to be runaways. 
\citet{ELT11} studied runaways assuming the binary supernova scenario, 
in which the companion star obtains a high velocity when one of the binary components 
undergoes a supernova explosion. They predicted about 0.5 per cent \footnote{\citet{ELT11} 
provides two values, 0.5 (mentioned in the text) and 2.2 per cent, 
for the fraction of O-star runaways produced by a supernova explosion of the primary star 
in a binary system. The former is obtained using O stars with velocity larger than 30~\kms, 
which is also our runaway criterion, the latter is derived with a lower velocity limit 
of $5~\kms$ for runaways. The latter value is applicable to process (iii) in the text.} 
of O stars being runaways when only this process is accounted for. The value combining ours and \citet{ELT11} 
is thus well consistent with the observed one. 

Figure~\ref{nmcl} demonstrates that the presence or absence of initial mass-segregation leads 
to differences in the numbers of ejected O stars only for lower mass clusters ($\mcl \lesssim 10^3~\msun$) 
with the constant half-mass radius we assume in this study. 
If larger $\rh$ are assumed the cluster mass at which the difference vanishes would move up to 
a larger cluster mass since the mass-segregation time-scale gets longer.  

\section{Properties of ejected O star systems}
In this section, we study how some properties, such as velocities, distances from the cluster centre, 
and system masses of the ejected O-star systems vary with cluster mass. We remind the reader that 
our analysis is restricted to systems younger than 3~Myr.

\subsection{Kinematics of the ejected O star systems}

\begin{figure*}
 \center
 \includegraphics[width=59mm]{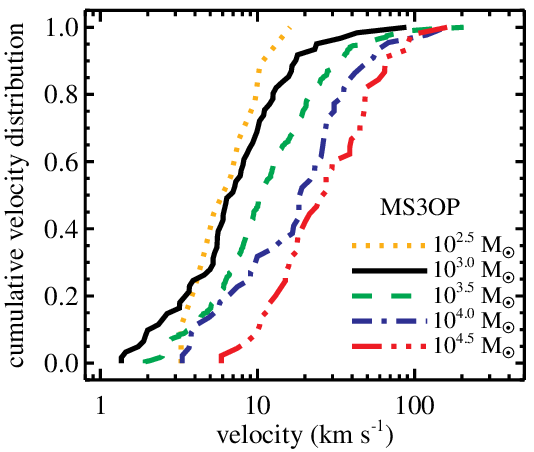}
 \includegraphics[width=59mm]{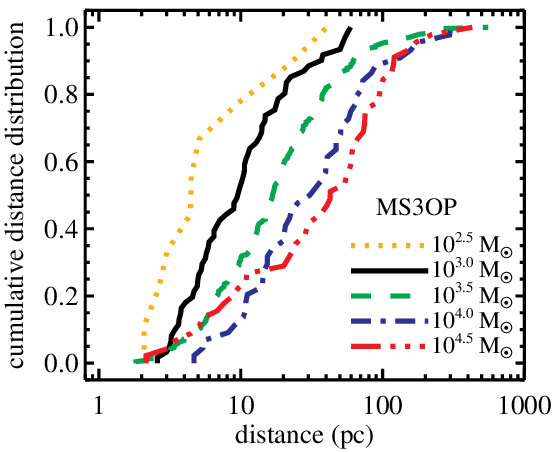}
 \includegraphics[width=59mm]{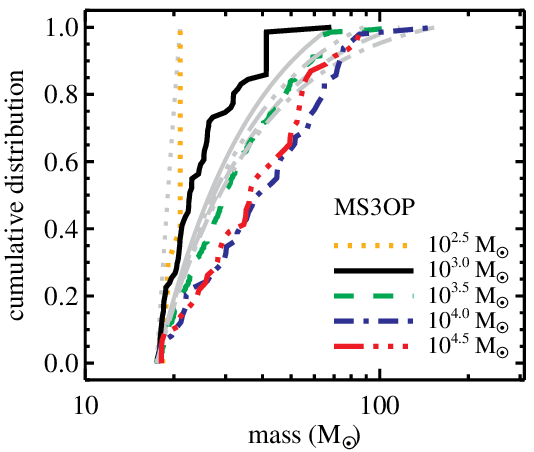}\\
 \includegraphics[width=59mm]{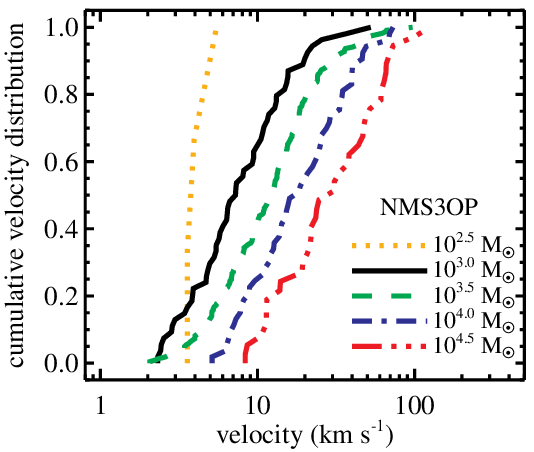}
 \includegraphics[width=59mm]{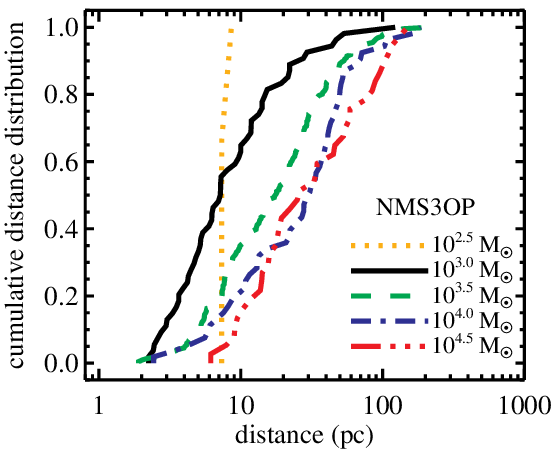}
 \includegraphics[width=59mm]{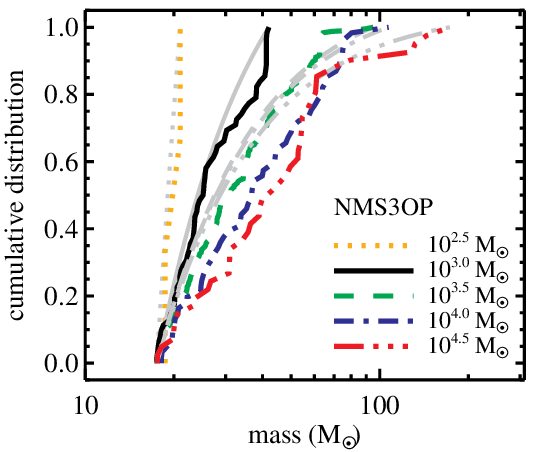}\\
 \includegraphics[width=59mm]{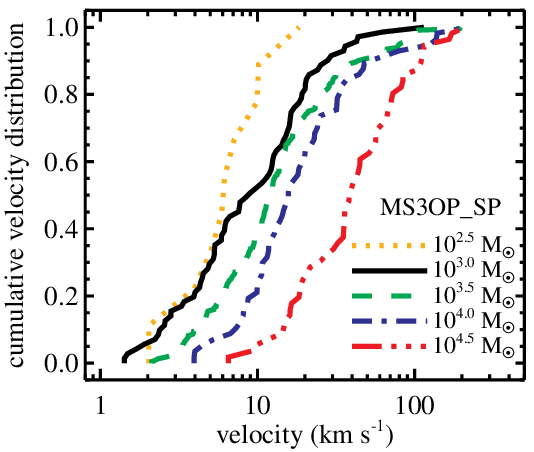}
 \includegraphics[width=59mm]{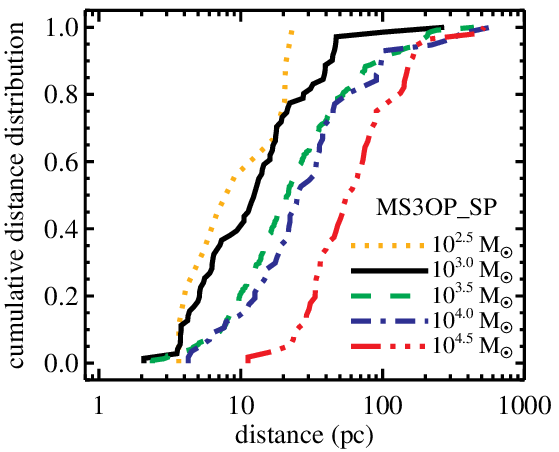}
 \includegraphics[width=59mm]{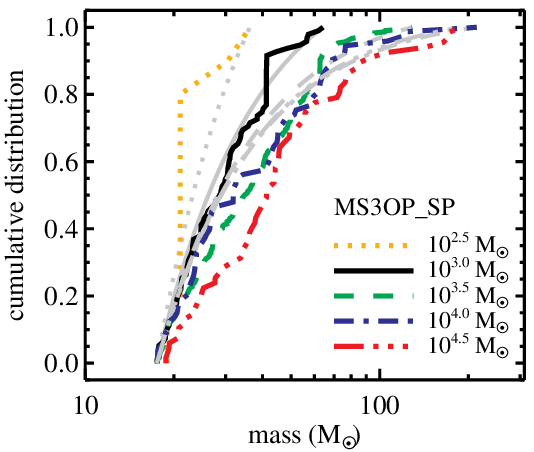}\\
 \includegraphics[width=59mm]{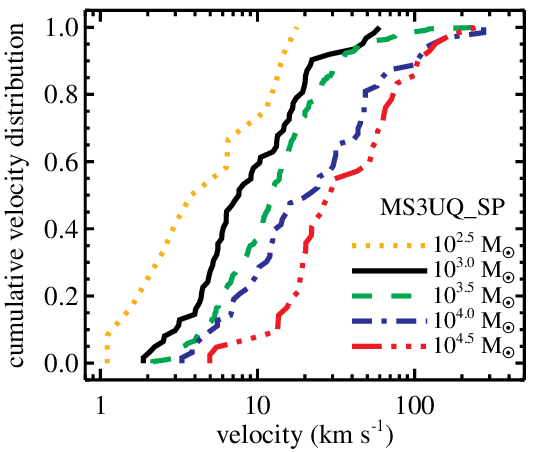}
 \includegraphics[width=59mm]{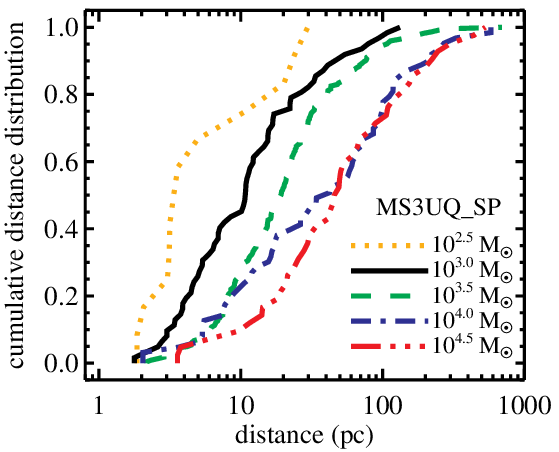}
 \includegraphics[width=59mm]{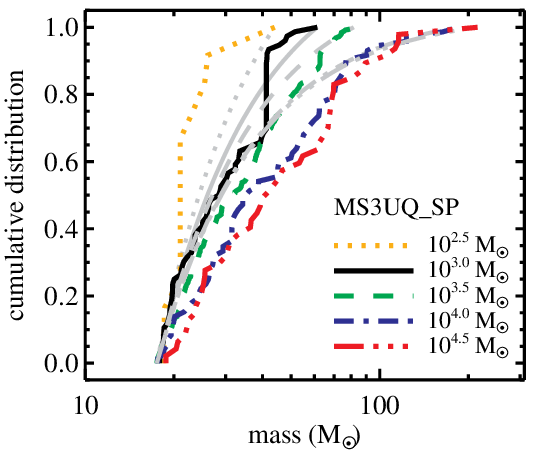}\\
  \caption{Normalised cumulative distributions of velocity ({\it left}) and of distance from the cluster centre ({\it centre})
 of ejected O star systems, and of mass of the individual ejected O stars ({\it right}) from our calculations. 
 See the text for the grey lines in the right most column.}
 \label{cdfig}
\end{figure*}
\setcounter{figure}{\value{figure}-1} 
\begin{figure*}
 \center
 \includegraphics[width=59mm]{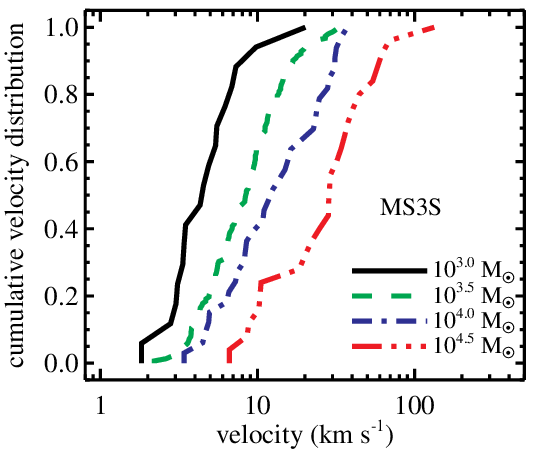}
 \includegraphics[width=59mm]{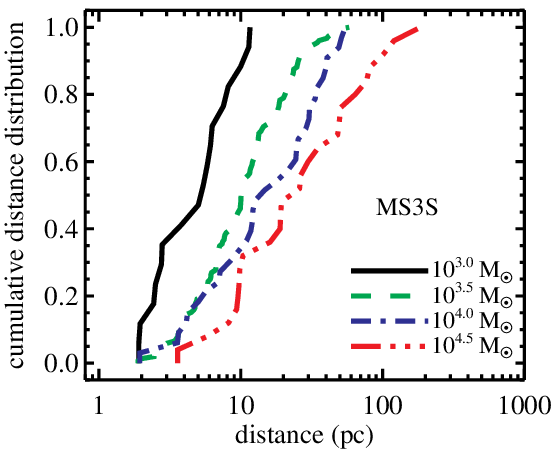}
 \includegraphics[width=59mm]{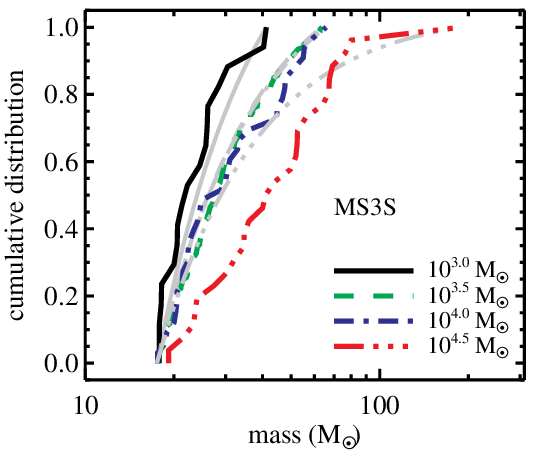}\\
 \includegraphics[width=59mm]{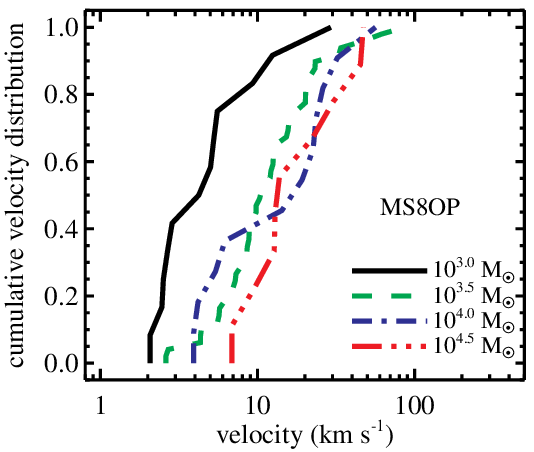}
 \includegraphics[width=59mm]{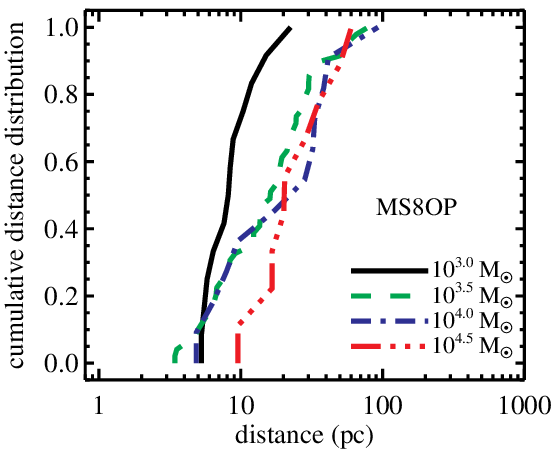}
 \includegraphics[width=59mm]{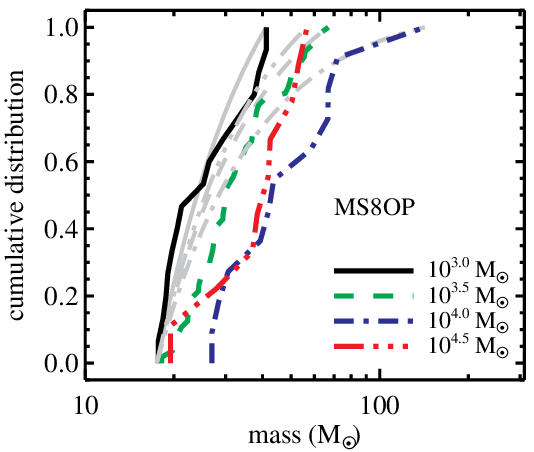}\\
 \includegraphics[width=59mm]{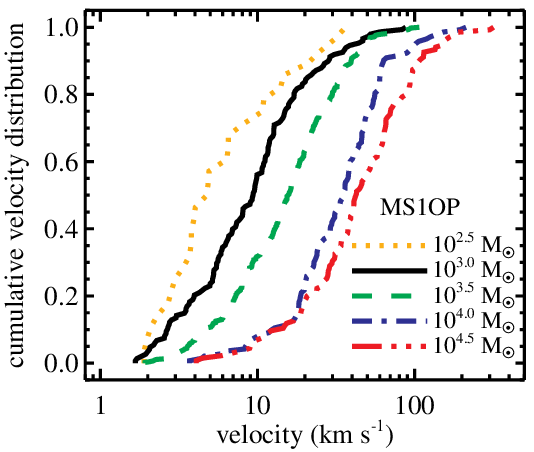}
 \includegraphics[width=59mm]{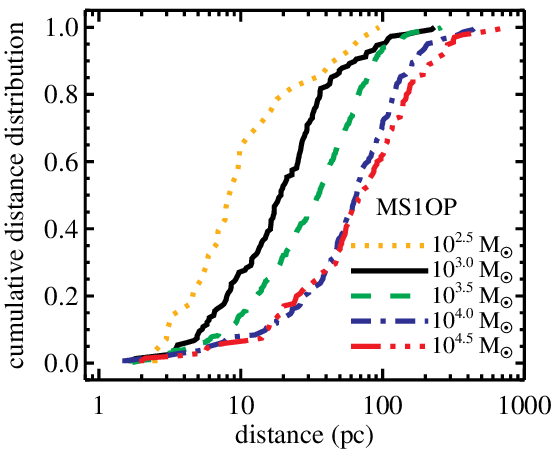}
 \includegraphics[width=59mm]{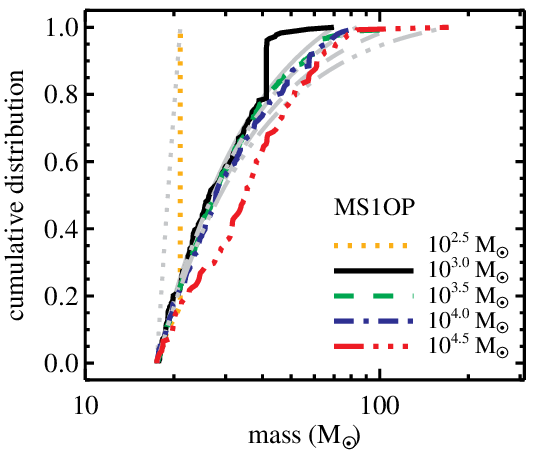}\\
 \caption{(Continued)}
\end{figure*}

Cumulative distributions of velocity, distance from the cluster centre, and system-mass 
of all ejected O-star systems for each cluster mass and for each set of cluster models are shown in Figure~\ref{cdfig}. 
Over all, the systems ejected from more massive clusters have velocity distributions skewed towards 
higher velocity. This can be understood in two ways. Firstly, the velocity dispersion of stars in a cluster 
increases with cluster mass  under the assumption of a constant cluster size 
(in general, velocity dispersion is a function of both size and mass of a cluster), 
i.e., on average stars in more massive clusters have higher velocities. 
Secondly, a star needs a velocity greater than the escape velocity which also 
increases with cluster mass, in order to leave its parental cluster.   
For the MS8OP model sequence (with $\rh=0.8$~pc), the trend is not as pronounced 
as in other models since the ejection 
fractions are small and so are the total number of ejected O-star systems, 
e.g., only in total 11 O stars are ejected out of 4 runs in the $10^{4.5}~\msun$ cluster models. 
Also in the case of the $10^{2.5}~\msun$ clusters in the NMS3OP models ($\rh=0.3$~pc) 
only three O star systems are ejected in 100 runs.   

The highest velocity of an ejected system generally increases with cluster mass, 
although this is not always the case, e.g, in the MS3OP model sequence, the system moving the fastest 
(a single star of mass $46~\msun$ with a velocity of $\approx200~\kms$) is ejected 
from a $10^{3.5}~\msun$ cluster. 
The velocity distribution is dependent on the cluster model, especially on the initial half-mass radius of the cluster.  
The MS1OP models have velocity distributions biased towards higher velocities 
compared to the larger-sized cluster models (Figure~\ref{cdfig}). 
The highest velocity of an ejected system (a single star of mass $34~\msun$ 
from an $\mcl=10^{4.5}~\msun$ cluster) is $\approx 300~\kms$ in the MS1OP model sequence.  

As the distance distribution of the ejected systems is directly related to the velocity distribution, 
their shapes are similar. The slight difference between them results from the time of ejections occurring 
at systematically different times in different models. 
Owing to the systematic shift to larger ejection velocities,  
the ejected systems travel on average further away from the clusters in the more massive clusters.  
The distances of the ejected systems from $\mcl\ge10^{4}~\msun$ clusters ranges from a few pc to a few 100 pc. 
The most extreme system (a single star with a mass of $34~\msun$) ejected from an MS1OP model traveled up to $\approx750$~pc away from its birth cluster by 3~Myr. 

Cumulative mass distributions of the ejected O stars are plotted in the right column of Figure~\ref{cdfig}. 
In the case of a binary system, masses of both components are counted separately. 
As binary fractions of the ejected systems are low for most of the models, 
the cumulative distributions of system masses would be similar to  those of individual stellar masses. 
A few models suggest that the more massive clusters may have a higher fraction of more-massive ejected stars. 
However the trend is not strongly evident. For example, in the case of the MS1OP models, the mass distributions 
of the ejected O stars look similar to each other for different cluster masses.
But only clusters with $\mcl\ge10^{3.5}~\msun$ eject O stars heavier than $100~\msun$  
as the lower mass clusters do not have such a massive star initially, 
e.g., for $\mcl=1000~\msun$, $\mmax\approx44~\msun$. About 15 per cent of the ejected O stars from clusters with 
$\mcl \gtrsim 10^{3.5}~\msun$ in the MS3OP\_SP models are more massive than $100~\msun$, 
while this is the case for $\lesssim 5$ per cent from the other models.
This may result from a higher number of mergers in the MS3OP\_SP models due to the high fraction of close, 
eccentric binaries in comparison to the other models.

Thin grey lines in the right column of Figure~\ref{cdfig} indicate cumulative mass distributions of O stars
from the canonical IMF with the lower- and the upper-mass limit of, respectively, $17.5~\msun$ 
and the mass of the most massive star among the ejected O star for each cluster-mass set. 
It follows that most models show that ejected stars have 
a shallower slope of the mass function than the canonical value (index of 2.3, see Equation~(\ref{imf})) 
as also shown in \citet{FP11} resembling the mass function in the cluster core.

\begin{figure}
 \center
 \includegraphics[width=85mm]{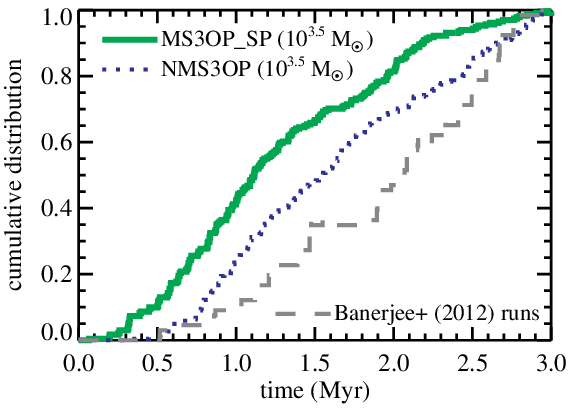}
 \caption{Cumulative O star ejection as a function of time for $10^{3.5}~\msun$ clusters 
   in models MS3OP\_SP and NMS3OP, and the \citet{BKO12} model. For example, almost 45 per cent 
   of all ejected O star systems are ejected by 1~Myr in the MS3OP\_SP ($10^{3.5}~\msun$) model. }
 \label{tcej}
\end{figure}

 The normalised cumulative number of the ejected O star-systems as a function of time 
for a few models are shown in Figure~\ref{tcej}. In the case of the MS3OP\_SP model ($\mcl=10^{3.5}~\msun$), 
almost 70 per cent of the ejected stars travel beyond 3 times the half-mass radius 
 before 1.5~Myr. Initially not mass-segregated cluster models 
(e.g., the $10^{3.5}~\msun$ clusters of the NMS3OP model sequence in Figure~\ref{tcej}) 
show a delay of the ejections as the model requires time for massive stars to migrate to the cluster centre 
to interact with other massive stars. But about half of the ejected stars already reach 
a distance of 3 times the half-mass radius from the cluster centre before 1.5~Myr of evolution. 
Also the ejection rate is higher during the earlier times of evolution. 
Note that the time at which a star is classified as being ejected is different to 
when the ejection has actually happened. The time presented in the figure is comprised of 
the traveling time the stars need to reach to the 
ejection criterion and thus the ejections actually happened at an earlier time 
than shown in the figure. This can naturally explain the steeper increase 
of the number of ejected stars at a later time for the \citet{BKO12} model 
in which the initial half-mass radius of the clusters is larger ($\rh=0.8$~pc) 
than in the above two models ($\rh=0.3$~pc). 
Thus an ejected star requires more time to reach the ejection criterion in their model 
(see Section~3 for the criteria). 
A rough estimate of the time when the ejections occur peaks at around 1~Myr, 
though it can vary with different initial conditions (Oh et al. in preparation).

\subsection{The binary fraction among the ejected O star systems} 
\begin{figure}
 \center
 \includegraphics[width=85mm]{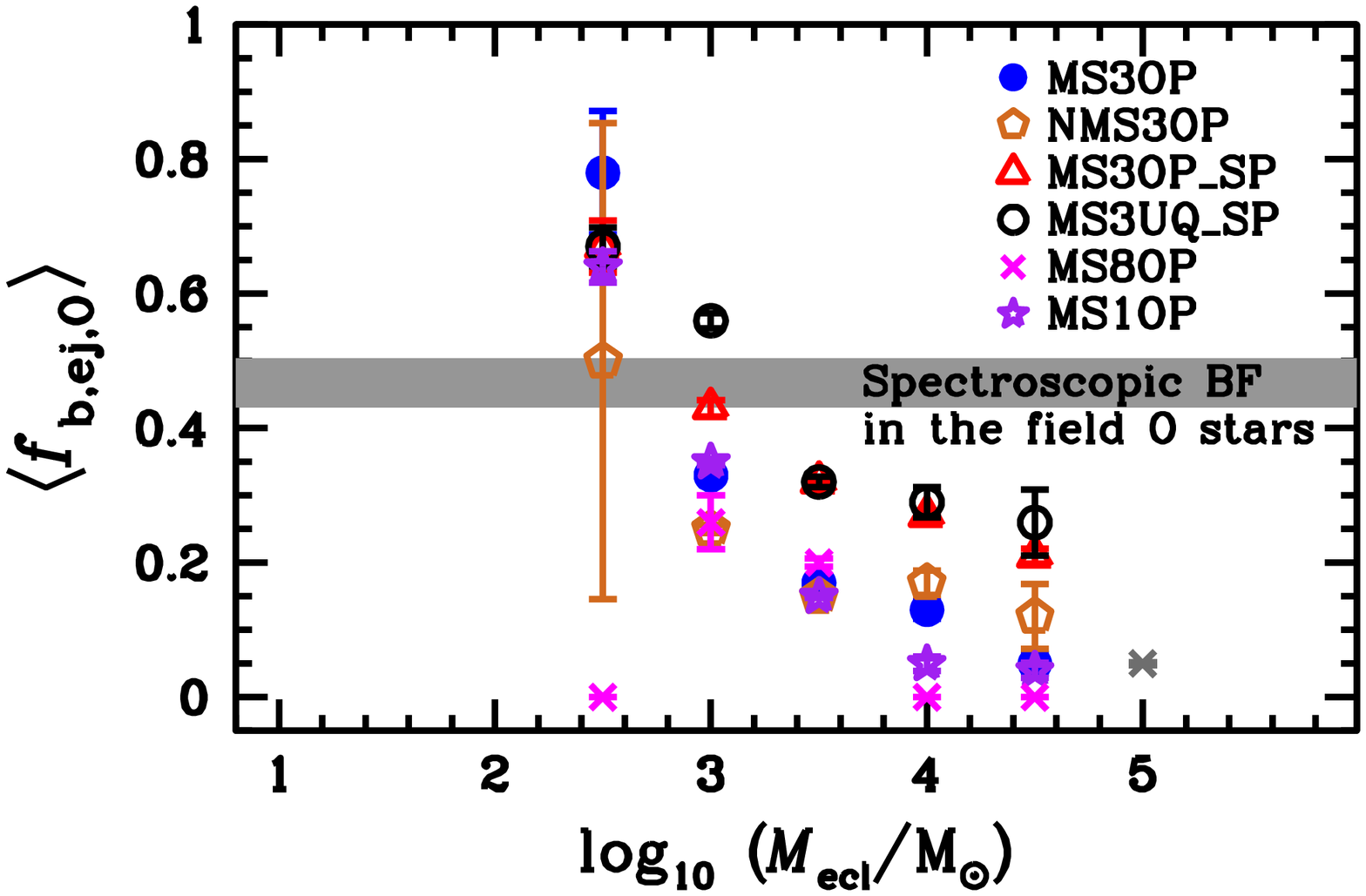}\\
 \includegraphics[width=85mm]{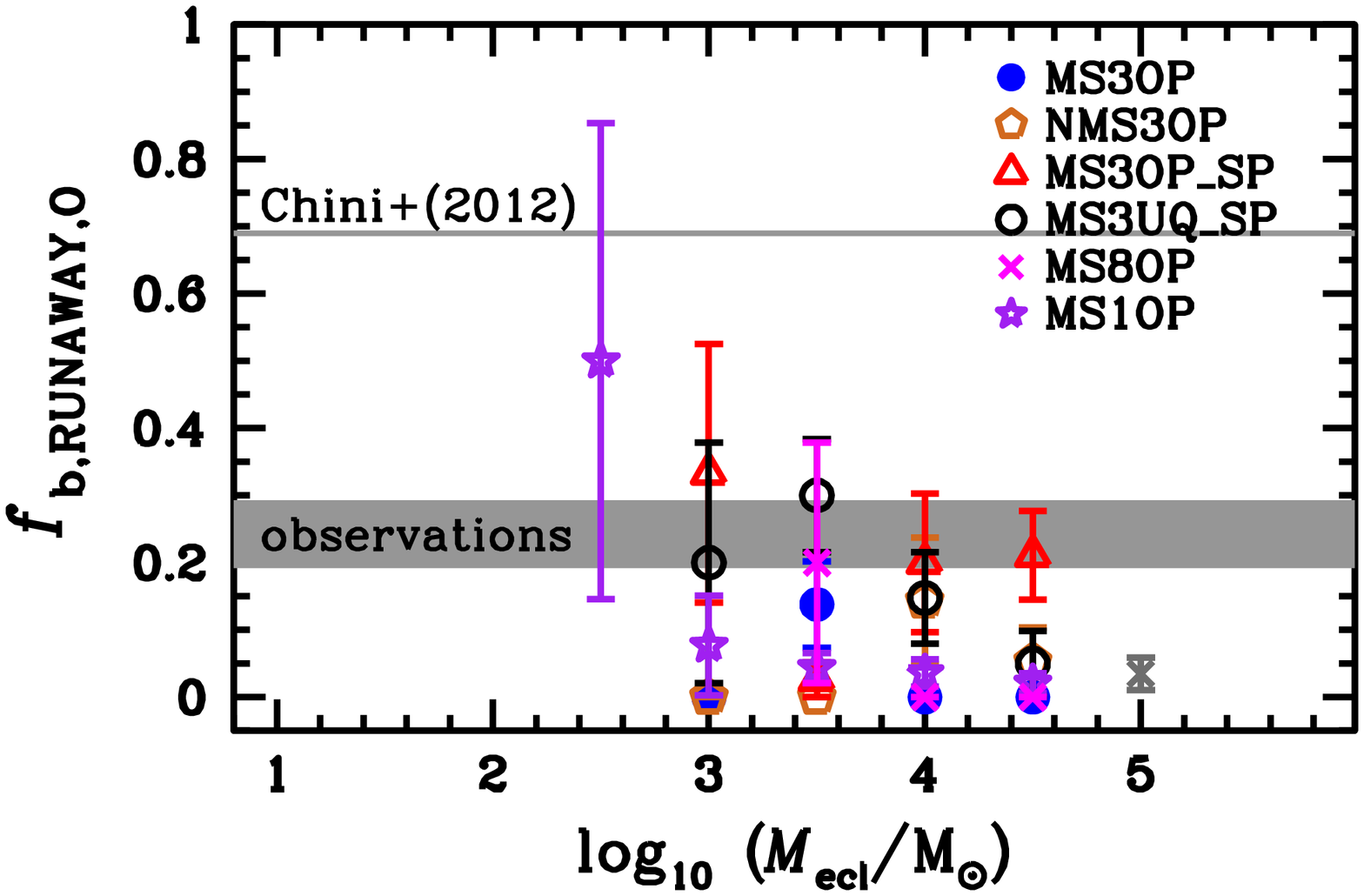}
 \caption{ Top: Averaged binary fraction among the ejected O-star systems,  
 $\langle f_{\mathrm{b,ej,O}}\rangle$, from initially binary-rich cluster models. 
 A (grey) cross at $\mcl=10^5~\msun$ is the \citet{BKO12} model which adopted a uniform initial period distribution for O star binaries. Error bars indicate standard deviations of $\langle f_{\mathrm{b,ej,O}}\rangle$. 
 The observed spectroscopic binary fractions in the Galactic field O stars are
shown as a grey area \citep{Gies87,Met98,Met09,Cet12}. 
 Bottom: Binary fraction among the runaway O stars from initially binary-rich cluster models. 
 The cluster models that do not eject any runaway O star, 
 e.g., $10^{2.5}~\msun$ cluster models from all sequences but the MS1OP sequence, are not included in the figure. Error bars indicate Poisson errors. The grey area indicates a range of the observed spectroscopic binary fractions of runaway O stars in \citet{Gies87} and \citet{Met98,Met09}. The grey horizontal line is the observed value in \citet{Cet12}.}
 \label{fejbin}
\end{figure}

Not only single O stars but binaries containing O star component/s are ejected from star clusters via dynamical interactions.
The binary fraction of ejected O stars can be substantial, up to $\approx78$ per cent (Table~\ref{tabresult}).
Even initially single-star clusters eject O star binaries dynamically (Table~\ref{tabresult}).  
In Figure~\ref{fejbin} is plotted the average binary fraction among ejected O-star systems 
from the binary-rich cluster models as a function of cluster mass.
A smaller fraction of binaries is ejected from the more massive clusters 
because the kick required to remove a system from the cluster is larger and more binaries are disrupted 
through interactions with other stars in the more massive clusters. 

The binary fraction among ejected O stars can be significantly altered by assuming a different initial period distribution. 
An initially higher fraction of close binaries leads to a higher binary fraction of ejected O stars.
Thus the more realistic MS3OP\_SP and MS3UQ\_SP models produce higher binary fractions of 
ejected O star systems compared to the other models (Figure~\ref{fejbin}). 
 The observed spectroscopic binary fraction among the Galactic field O stars 
is 43 -- 50 per cent \citep{Gies87,Met98,Met09,Cet12}. This may be reproduced readily 
through dynamical ejections of O stars with more-realistic 
\citep[than the][]{PK95b} period distribution function for O star binaries, 
such as constrained recently by \citet{Set12} (Equation~(\ref{SP})).  
This comes about because, as can be seen for the most realistic sequence (MS3UQ\_SP), 
the dominant contributors to the population of ejected O star systems are 
$\le 10^4~\msun$ clusters (lower panel of Figure~\ref{nmcl}), 
and these ejected O star systems have a binary fraction $\gtrsim30$~per~cent 
(upper panel of Figure~\ref{fejbin}).  It should be noted that 
the given observed binary fractions are lower limits of the binary fraction 
of the field O stars since only spectroscopic binaries are counted. 

In the case of runaways, the binary fractions from our most realistic sequences, 
MS3OP\_SP and MS3UQ\_SP, are well consistent with the observed spectroscopic binary fraction 
of the runaway O stars in \citet{Gies87} and \citet{Met98,Met09}, 
19 -- 29 per cent (grey area in the lower panel of Figure~\ref{fejbin}), but too small compared 
to the value in \citet{Cet12},\footnote{We note here that the \citet{Cet12} sample 
is not biased. They selected stars from the Galactic O Star Catalogue v.2.0 \citep{Sota08} of which 
248 stars could be reached from their observatory. They additionally employed archival data 
from the European Southern Observatory (ESO) for those stars already contained in their survey. 
Thus the ESO spectra did not add any new star to their sample but increased only the number 
of spectra and, as a consequence, the time basis for individual stars (R. Chini, private communication).} 
69 per cent (grey horizontal line in the same figure).
This may be due to the difference in assigning runaways between the \citet{Cet12} work and the other studies. In \citet{Cet12}, stars are classified as runaways when they can be traced back to any cluster or associations in the study of \citet{SR08}. In other words, the runaways in their sample may be simply ejected O stars from star clusters which have masses $\lesssim 10^{4}~\msun$ since very massive clusters
are exceedingly rare. The other studies assign stars to runaways with a large distance from the Galactic plane or with a high peculiar (radial) velocity.

In most of our cluster models (with the exception of the MS3OP\_SP and MS3UQ\_SP models) 
the binary fraction of O stars, whether they are inside a cluster or not, at 3~Myr is 
smaller than the observed O star binary fraction \citep[$\approx 70$ per cent,][and references therein]{Cet12,Set12}.
This is the case because dynamical interactions between binaries and other cluster members
disrupt binaries, preferentially those with a wide orbit \citep{PK95a,Pet09,MKO11} and 
because the Kroupa period distribution 
(which was constrained originally only for late-type stars) extrapolated 
in some of the model sequences to the O star regime has a larger fraction of long period binaries 
compared to the observed distribution of O star binaries (Figure~\ref{pcdf}).

\section{Discussion}
While the mass of clusters which eject O stars most efficiently is the same for all models ($\mcl\approx10^{3.5}~\msun$), 
it has been shown that the dynamical ejection of O stars can be highly sensitive to the initial 
conditions of star clusters. In this section, we further discuss the effects of 
our choice of initial parameters for massive stars and for the clusters, and the available observational constraints.

The ejection fraction is strongly dependent on the initial cluster radius. As seen in Figure~\ref{figfej}
the shapes of the curves look similar but the values of the ejection fraction are larger for initially smaller sized clusters. 
Thus it is important to check if our models are comparable to observed young star clusters. 
The half-mass densities of our \nbody\ models at 3~Myr 
range from $175$ to $\approx10^{5}~\uden$, from $\approx 160$ to $\approx 3.4\times10^{4}~\uden$, 
and from $\approx 50$ to $\approx4000~\uden$ for initial $\rh=0.1$, $0.3$, 
and $0.8$~pc clusters, respectively. 
These values are well within the densities of observed young ($<5$~Myr) massive 
($\mcl > 6000~\msun$) star clusters, classified as star burst clusters in \citet{PS09}, 
which range from 1500 to $4\times10^{5}~\uden$ \citep{FD08}. 
 However, caution is due in directly comparing the model cluster densities 
to the observed values since the observed densities are closer to the central (i.e., highest) 
densities rather than to the half-mass density. 
Concerning the initial radii of the embedded clusters, these are dynamically sub-pc 
 \citep{Tet99,LL03,KP05,MK11,SSB05,Tet09,Tet11,Tet14}.

The initial binary population plays a major role in the dynamical ejections of massive stars 
(see the differences between the MS3OP and MS3S models). Thus it is important to adopt realistic initial 
binary parameters to understand massive star ejections in reality. 
The pairing method for massive binaries adopted in most of our models results in extreme mass-ratios, 
e.g., most of them are larger than 0.9. This was motivated by the high fraction of twins in observed massive binaries 
\citep[][ but see also \citealt{LL06}]{PS06,KF07}, 
while recent observational studies favour a rather flat mass-ratio distribution for O-type binaries 
\citep{KK12,Set12} still excluding random pairing from the IMF for this spectral type. 
Adopting a more realistic initial mass-ratio distribution for O-type binaries may reduce 
the efficiency of the ejection, though the shape of the $\fej(\mcl)$ ejection fraction as a function of 
parent birth-cluster mass, is expected to be kept. 
In this contribution we include a realistic model for the observed period distribution of O star binaries (Equation~(\ref{SP})). 
The model with the observed period distribution, which reproduces the observed abundance of short period binaries, exhibits 
a substantial increase of the ejection fraction, by $\approx30$ per cent, from massive clusters 
($\mcl\ge 10^{4}~\msun$).  Whether the observed distribution is a real initial distribution 
is still unclear as the observed distribution has been compiled from several clusters 
which may have already evolved dynamically despite their young ages (1--4~Myr).  
The authors suggested that dynamical evolution would be negligible for such massive close binaries 
($\log_{10}(P/\mathrm{days})<3.5$). However, the population of massive binaries can evolve through the dynamical interactions 
with other massive stars/binaries in the clusters (Oh et al. in preparation).  Moreover, short-period massive binaries 
are vulnerable for merging either due to stellar evolution or the perturbation-induced collision of the two binary-star companions. 
Thus the effect of dynamical evolution needs to be taken in to account iteratively to find 
the true initial period distribution of massive binaries. 
But this is beyond the aim of this study. For late-type binaries this has already been performed successfully by \citet[][see also \citealt{MKO11,MK12}]{PK95a,PK95b}.

Unlike the other initial conditions mentioned above, initial mass segregation does not result in 
a difference of the O star ejection fraction for clusters with $\mcl\ge10^{3}~\msun$ 
and initially $\rh=0.3$~pc due to efficient dynamical mass segregation. 
But differences do occur for lowest mass clusters: initially mass-segregated clusters more efficiently eject
O stars than non-segregated clusters, as dynamical mass segregation is inefficient to gather massive stars 
in the core of such clusters.
For clusters with larger initial $\rh$ values, as the dynamical mass segregation time scale becomes longer, 
the initial mass segregation can play a significant role. 
In observations, post gas-expulsion young massive clusters (age $\lesssim5$~Myr) are generally compact 
\citep[effective radii $\lesssim$ a few~pc,][]{PMG10,KP05,Tet03}.

 Throughout this study the $\mmax$--$\mcl$ relation is adopted.  
 It follows from an intuitive theoretical concept that the mass of a star forming
molecular cloud core is finite and that star formation is a self-regulated process \citep{AF96}, and
it follows from the data on young stellar populations \citep{WKB10,WKP13,KM11}. The existence of 
a physical $\mmax$--$\mcl$ relation is thus taken to be established \citep{WKP14}.
Although the theory has been successful in describing the stellar contents of whole galaxies \citep[e.g.,][]{Wet13}, 
it has been debated if the relation exists \citep[see section~2.1 in][for details]{WKP13} or 
if star formation is a purely stochastic process \citep[e.g.,][]{KM15}. 
This is a fundamental issue for understanding galaxy-wide IMFs and how massive stars form, 
and  we here thus stress that the data do not support the hypothesis that star formation is stochastic \citep[see section~7 of][]{KP14}. 
For example, \citet{Het12,Het13} presented that L1641, the low-density star-forming region 
(containing as many as $\approx 1600$ stars) of the Orion A cloud, is deficient of massive 
(O and early B) stars compared to the canonical IMF with 3--4$\sigma$ significance and that with a probability of only 3 per cent
the southern region of L1461 and the ONC can be drawn from the same population 
supporting that the high-mass end of the IMF is dependent on environmental density. 
 In section 4.2 of \citet{KM15} three references \citep{Calzetti10, Fumagalli11, Andrews13} are cited as containing evidence against a physical $\mmax$--$\mcl$ relation. \citet{WKP14} critically discuss these works showing \citet{Andrews13} data to be entirely  consistent with the physical $\mmax$--$\mcl$ relation. 
As already discussed in Section~3, whether the $\mmax$--$\mcl$ relation is adopted or 
not does not affect the here obtained results though.  

We do not include a gas potential, which would subsequently be removed rapidly, 
in the calculation \citep[cf.,][]{KAH01,BK14}. 
The main outcomes of gas expulsion are significant stellar-loss 
from and expansion of clusters and their possible subsequent dissolution. 
Therefore gas expulsion can weaken the ejection efficiency by lowering the 
(central) density of a cluster if that occurs at a similar time 
when a majority of the ejection processes take place. 
In the future, further studies including a gas potential are needed to understand the effect of 
gas expulsion on the ejection of massive stars. Inclusion of a gas potential will be possible 
once the statistical quantification of the gas expulsion process is in place and computable 
\citep[cf.,][]{BK14}. It has been argued recently that gas expulsion, if the process occurs at all, 
may have no effect on star cluster evolution \citep[e.g., see][and references therein]{Let14}. 
It may, however, be difficult to detect the expanding signature resulting from residual gas expulsion since 
revirialization after gas expulsion can be rapid for massive clusters, $\mcl>10^3~\msun$ \citep{PMG10,BK13}.

To constrain which initial conditions are responsible for the observed field O star populations, 
one needs the observed quantities to compare with the results from models.  
However, it is very difficult to find which clusters individual field O stars originate from. 
Some ejected O stars can travel further than a few hundred pc away from their birth cluster 
within a few Myr thus, without knowledge of their full-3D kinematics and the Galactic potential, 
it is almost impossible to find their birth cluster.
\facility{Gaia} may help to find the birth clusters of the field O stars. 
However, the two-step ejection process of \citet{PK10} complicates this.
 Their calculation suggests that 1--4.6 per cent of O 
stars would appear to have formed in isolation. They used the observed values 
of the O star runaway fraction (10--25 per cent, \citealt{GB86}; 46 per cent, \citealt{Stone91}),  
and the binary fraction among the runaways \citep[10 per cent,][]{GB86}. 
 Since  most, if not all, ejected O-star--O-star binaries may 
go through the two-step ejection process and then cannot be traced back to their birth cluster, 
the fraction of O stars which appear to have formed in isolation can be higher 
if a binary fraction of the dynamically ejected stars is substantial. 
 A quantitative study is required to determine the fraction of O-stars experiencing the two-step ejection process.
 
\section{Summary}
We study how the ejection of O-stars varies with cluster mass (or density 
as we assume a constant size for different cluster masses) under diverse initial conditions.
We find that clusters with a mass of $10^{3.5}~\msun$  most efficiently shoot out
their O stars to the field compared to other ones with other masses, for all models, 
i.e., this result obtains independently of radius or the presence of initial mass segregation or binaries.  
Our results suggest that moderately massive clusters ($\mcl=10^{3.5}~\msun$), which have formed 
about 10 O star systems, most efficiently eject their O stars  
through energetic encounters between massive stars in the cluster core. 
Up to on average $\approx 25$ per cent of the initial O star content is ejected 
from the clusters with this cluster mass in $\rh=0.3$~pc models 
(52 per cent in $\rh=0.1$~pc clusters, Figure~\ref{figfej}). 
However, the spread is large and ejections of all O stars leading to 
remnant young clusters void of O stars occur in about 1 per cent of all such clusters (Figure~\ref{figfej}). 
\citet{FP11} suggested that one bully binary, dynamically formed, is  mainly responsible for flinging out 
the stars from initially single-star clusters. Here we show that the number of ejections is significantly higher 
when the highly significant initial massive binary population is accounted for. 
The ejections, thus, occur via close encounters involving several binaries in reality 
since the massive stars are observed to have high multiplicity 
fractions in young star clusters \citep[and references therein]{Cet12,Set12}. 
Furthermore, more realistic clusters, which have O-star binaries with periods following the \citet{Set12} 
distribution (MS3OP\_SP), 
lead to even larger ejection fractions, especially for massive clusters. 
The mass-position of the peak of $\fej=\fej(\mcl)$, however, does not change 
($M_{\mathrm{ecl,peak}}\approx10^{3.5}~\msun$).

The decrease of the O star ejection fraction with increasing cluster mass for massive 
($\mcl \ge 10^{3.5}~\msun$) clusters can be represented by a simple power-law of cluster mass 
with an exponent between $-0.16$ and $-0.43$.  
Clusters with an initial half-mass radius $\rh=0.3$~pc and with relatively simple initial conditions 
(Kroupa initial period distribution for O stars or no initial binaries) 
have an exponent of $\approx -0.4$ being in good agreement with $-0.5$ which is derived from 
the number of O stars from the IMF and the binary-single star scattering rate (Equation~(\ref{eqfej})). 
Models with a high fraction of close initial binaries or smaller $\rh$ deviate from 
the expected theoretical value by having larger ejection fractions.   

We show that not only the ejection fraction but also the properties of the ejected O star systems 
depend on the initial cluster mass. The velocity distribution of 
the ejected O star systems shifts to higher velocities for more massive clusters, and so does 
their distance distribution from the cluster. But the distribution of ejected system masses 
varies weakly with cluster mass. 
The binary fraction of the ejected systems decreases with increasing cluster mass 
as does the binary fraction of the systems remaining in the cluster because binary disruption 
via interactions with other members of the cluster is more efficient for a denser cluster, 
i.e., more massive cluster. The binary fraction among the ejected O star systems is substantial, 
especially in the more realistic models in which the period distribution of 
the observed O star binaries is implemented (MS3OP\_SP and MS3UQ\_SP). 

Considering that the cluster mass function is approximately a power-law with a near Salpeter index, 
the main fraction of field O stars dynamically ejected from clusters originate 
from low/intermediate mass clusters ($\mcl\lesssim$ a few $1000~\msun$, see Section~5). 
 The dynamical ejection process populates $\approx16$ to $38$ per cent of all O stars, 
formed in clusters, to the field. Combining our results on the dynamical ejection fraction 
and the binary-supernova scenario computed by \citet{ELT11} 
 it follows that the observed fractions of the Galactic field- and runaway O stars are well
accounted for theoretically. 
It should be noted that very massive ($\gtrsim100~\msun$) runaways can only originate from 
massive star-burst clusters ($\mcl\gtrsim 10^{4}~\msun$) because only massive clusters 
can harbour a sufficient number of very massive stars to eject some of them.

\acknowledgments
We are grateful to Sverre Aarseth for making his NBODY6 code freely available and for continuing its improvements. 
We also thank Rolf Chini for comments on the observational data in his paper.

\begin{appendix}
\counterwithin{table}{section}
\renewcommand{\thetable}{\Alph{section}\arabic{table}}
\section{A Table for the results of individual models}
 Here we append the table containing the results and properties of each model cluster (Table~\ref{tabresult}).
\begin{deluxetable*}{lcccccccc}
\tabletypesize{\footnotesize}
\tablewidth{0pt}
\tablecaption{Results of all models at 3~Myr.\label{tabresult}}
\tablehead{
\colhead{$\mcl$ ($\msun$)} &\colhead{$\mmax$ ($\msun$)} &\colhead{$\langle \rh \rangle$ (pc)}
&\colhead{$\anos$} &\colhead{$\anej$} & \colhead{$\afej$} 
&\colhead{$\langle f_{\mathrm{b,O}}\rangle$} & \colhead{$\langle f_{\mathrm{b,ej,O}}\rangle$}
&\colhead{$N_{\mathrm{run}}$}}
\startdata
\sidehead{MS3OP}
$10^{2.5}$  &21.2   &0.60 &1.11     &0.09   &0.080 &0.84  &0.78 &100\\
$10^{3}  $  &43.9   &0.72 &2.96     &0.61   &0.172 &0.66  &0.33 &100\\
$10^{3.5}$  &79.2   &0.70 &9.57     &2.38   &0.252 &0.45  &0.17 &100\\
$10^{4}  $  &114.7  &0.50 &30.10    &4.40   &0.149 &0.23  &0.13 &10 \\
$10^{4.5}$  &136.2  &0.50 &105.00   &11.25  &0.105 &0.15  &0.05 &4  \\
\tableline
\sidehead{NMS3OP}
$10^{2.5}$ &21.2  &0.56 &1.14   &0.03 &0.020 &0.87 &0.50 &100\\
$10^{3}  $ &43.9  &0.68 &3.11   &0.54 &0.166 &0.70 &0.25 &100\\
$10^{3.5}$ &79.2  &0.65 &9.26   &2.21 &0.244 &0.51 &0.15 &100\\
$10^{4}  $ &114.7 &0.57 &30.00  &5.30 &0.177 &0.32 &0.17 &10 \\
$10^{4.5}$ &136.2 &0.50 &98.25  &9.25 &0.092 &0.27 &0.12 &4  \\
\tableline
\sidehead{MS3OP\_SP}
$10^{2.5}$ &21.2  &0.62 &1.23   &0.09   &0.090 &0.79 &0.67 &100\\
$10^{3}  $ &43.9  &0.74 &2.92   &0.71   &0.228 &0.65 &0.43 &100\\
$10^{3.5}$ &79.2  &0.72 &9.11   &2.32   &0.257 &0.51 &0.32 &100\\
$10^{4}  $ &114.7 &0.66 &26.40  &5.70   &0.218 &0.40 &0.27 &10 \\
$10^{4.5}$ &136.2 &0.55 &96.50  &14.00  &0.145 &0.27 &0.21 &4  \\       
\tableline
\sidehead{MS3UQ\_SP}
$10^{2.5}$ &21.2  &0.58 &1.34    &0.12   &0.097 &0.86 &0.67 &100\\
$10^{3}  $ &43.9  &0.73 &3.32    &0.62   &0.177 &0.66 &0.56 &100\\
$10^{3.5}$ &79.2  &0.69 &10.28   &2.23   &0.209 &0.59 &0.32 &100\\
$10^{4}  $ &114.7 &0.63 &33.60   &6.30   &0.188 &0.45 &0.29 &10 \\
$10^{4.5}$ &136.2 &0.54 &103.00  &10.50  &0.104 &0.36 &0.26 &4  \\
\tableline
\sidehead{MS3S}
$10^{2.5}$ &21.2  &0.44 &1.17   &0.00 &0.000 &0.84 &0.00 &100\\
$10^{3}  $ &43.9  &0.55 &2.88   &0.17 &0.043 &0.56 &0.13 &100\\
$10^{3.5}$ &79.2  &0.59 &10.40  &1.52 &0.145 &0.22 &0.04 &100\\
$10^{4}  $ &114.7 &0.52 &34.30  &3.30 &0.096 &0.06 &0.00 &10 \\
$10^{4.5}$ &136.2 &0.48 &114.50 &6.25 &0.054 &0.02 &0.04 &4  \\
\tableline
\sidehead{MS8OP}
$10^{2.5}$ &21.2  &0.93 &1.10   &0.00  &0.000 & 0.97 &0.00 &100\\
$10^{3}  $ &43.9  &1.02 &2.45   &0.12  &0.036 & 0.92 &0.26 &100\\
$10^{3.5}$ &79.2  &1.05 &8.81   &0.49  &0.053 & 0.54 &0.20 &100\\
$10^{4}  $ &114.7 &1.02 &27.90  &1.10  &0.038 & 0.39 &0.00 &10 \\
$10^{4.5}$ &136.2 &1.00 &86.50  &2.25  &0.026 & 0.35 &0.00 &4  \\
$10^{5}\tablenotemark{a}$ &145.3 &0.90 &207.25 &17.50 &0.084 &0.62  &0.05& 4\\
\tableline
\sidehead{MS1OP}
$10^{2.5}$ &21.2  &0.60 &1.16   &0.28   &0.248 & 0.91 &0.64 &100\\
$10^{3}  $ &43.9  &0.63 &3.09   &1.48   &0.415 & 0.72 &0.35 &100\\
$10^{3.5}$ &79.2  &0.49 &9.10   &4.55   &0.516 & 0.47 &0.15 &100\\
$10^{4}  $ &114.7 &0.40 &30.90  &14.30  &0.468 & 0.22 &0.05 &10 \\
$10^{4.5}$ &136.2 &0.33 &106.00 & 33.75 &0.320 & 0.12 &0.04 &4                    
\enddata
\tablecomments{Columns~1 and 2 show the initial stellar mass of the model clusters 
 and the initial maximum stellar mass, respectively.
 Columns~3 and 4 present the averaged half-mass radius,$\langle \rh \rangle$, 
 and the average number of O systems present in the calculations, $\anos$, at 3~Myr, respectively. 
 Columns~5 and 6 are the average number of ejected O-star systems, $\anej$,
  and the average O-star system ejection fraction, $\afej$.
 The average binary fraction of O-star systems remaining in the cluster, $\langle f_{\mathrm{b,O}}\rangle$, 
 and the average binary fraction of ejected O-star systems, $\langle f_{\mathrm{b,ej,O}}\rangle$, at 3~Myr 
 are listed in columns~7 and 8, respectively.
 Note that the fraction of binaries among all O-star systems is smaller than the initial value of
 $f_{\mathrm{b,O}} = 1$ because systems are disrupted in these models.
 The number of realisations, $N_{\mathrm{run}}$, for each cluster mass is listed in the last column. 
 Only clusters with $\mcl \geq 10^{2.5}~\msun$ are listed as less massive clusters do not have O stars in our models. }
\tablenotetext{a}{The $10^{5}~\msun$ cluster model is adopted from \citet{BKO12}.
 The initial conditions of the model from \citet{BKO12} are slightly
 different from ours. Their O-star binaries are mostly close binaries (see Figure~\ref{pcdf}) 
 and this results in higher ejection fractions and 
 less dynamical disruptions of O-star binaries (higher binary fraction of O-star systems at 3~Myr) 
 compared to our models.
}
\end{deluxetable*}
\end{appendix}

\end{document}